\documentclass{aa}
\usepackage{txfonts}
\usepackage{graphics}
\usepackage{graphicx}	
\usepackage{amssymb}
\usepackage{afterpage}
\usepackage{natbib}
\begin{document}

\def\deg{$^{\rm o}$}
\def\ico{I$_{\rm CO}$}
\def\i14{I$_{\rm 1.4}$}
\def\Qco{Q$_{\rm CO/RC}$}
\def\qco{q$_{\rm CO/RC}$}
\def\qfir{q$_{\rm FIR/RC}$}
\def\matteo{MHE}
\def\io{PMH}
\def\P{$\bar{\rm P}$}
\def\n0{N$_{0}$}
\def\tc{$t_{\rm c}$}
\def\tsyn{$t_{\rm syn}$}

\defcitealias{matteo05}{MHE}
\defcitealias{io06}{PMH}

\title{Correlation of the radio continuum, infrared, and CO molecular emissions in \object{NGC\,3627} 
\thanks{Based on 
observations carried out with the IRAM Plateau de Bure Interferometer. IRAM is supported 
by INSU/CNRS (France), MPG (Germany) and IGN (Spain).}
}

\author{ R. Paladino \inst{1} \and M. Murgia \inst{1,2} \and A. Tarchi 
\inst{1,2} \and L. Moscadelli \inst{3} \and C. Comito \inst{4}}
\offprints{R. Paladino, rpaladin@ca.astro.it}

\institute{
INAF\,-\,Osservatorio Astronomico di Cagliari, Loc. Poggio dei Pini, Strada 54,
I-09012 Capoterra (CA), Italy
\and 
INAF\,-\,Istituto di Radioastronomia, Via Gobetti 101, I-40129 Bologna, Italy
\and 
INAF\,-\,Osservatorio Astrofisico di Arcetri, Largo E. Fermi, n. 5, I-50125, 
Firenze, Italy 
\and
 MPI, Bonn. Max-Planck-Institut f\"uer Radioastronomie, Auf dem H\"uegel 69, 
D-53121 Bonn, Germany
}

\date{Received; Accepted}

\abstract{}{ {We present new radio observations of two regions 
of the spiral galaxy \object{NGC\,3627}, including new radio continuum observations 
at 1.4 GHz with the Very Large Array, and also new observations 
in the CO line, taken with the Plateau 
de Bure interferometer.
Comparing these observations with archival {\it Spitzer} and H$_{\alpha}$ data we  
 study the correlation of the radio continuum (RC), infrared-8$\mu m$, and CO 
emissions at a spatial resolution of 100 pc.
}
 }
{We compare the point-by-point variations of the RC, CO, and 8$\mu m$ 
brightnesses in two 
distinct  regions of 2 kpc $\times$ 2 kpc in size of \object{NGC\,3627}. 
We also present a three-dimensional fit of the observed data. 
}
{We examined scale much lower 
than the electron diffusion scale, where a breakdown of the correlations
would be expected. However, no evidence for 
such correlation breakdown has been found. The RC emission follows 
the distribution of CO well, and the widths of several 
structures, measured along slices across them, are comparable. 
Furthermore, we found that down to a spatial scale of 100 pc,
 the radio continuum emission is correlated with dust emissions at 4.5, 5.8, and 
8 $\mu$m, which trace different dust temperatures. 
 We present a new perspective, a three-dimensional representation,
 with which to visualize and study the 
RC-CO-24$\mu m$ and 
RC-CO-8$\mu m$ correlations.
We fit the observed data with a three-dimensional line,  
obtaining a rms of 0.25 dex. 
} 
{ The observed correlation  enhances the complexity of the electrons diffusion, 
losses, and injection mechanisms, and of their connection to star formation 
processes described by molecular and dust emissions.
We plan to further investigate this connection using spatially resolved spectral 
index studies at low radio frequencies where the thermal emission component is 
seemingly negligible.
}

\maketitle

\section{Introduction}

The star formation process powers the emissions of spiral galaxies across  
the whole electromagnetic spectrum.
Star-forming galaxies are luminous sources of X-ray emission due to 
close accreting binaries with a compact companion, young supernova remnants, and hot 
plasma associated to star-forming regions and galactic winds \citep{fabbiano89}. 
Young massive stars are strong sources 
of ultraviolet (UV) and, via the ionization of the interstellar medium (ISM),
of H$_{\alpha}$ emissions \citep{rownd99}.
The dust heated by the interstellar radiation field produces 
the far infrared (FIR) luminosity. 
At millimeter wavelengths the CO emission traces the bulk of molecular  
gas in which star formation occurs \cite[e.g.,][]{YouSco91}, 
the HCN emission is also a reliable tracer of 
dense molecular regions.
The radio emission, at centimeter wavelengths, from galaxies consists 
of a mixture of non-thermal and thermal 
component, produced  via synchrotron emission and free-free emission from HII regions, 
respectively.
Correlations between global luminosities, integrated over the entire galaxies, 
at these different
 wavelengths have been reported by several authors. In particular, 
 \cite{ranalli03} studied a X-ray/radio correlation in star-forming galaxies;
  \cite{kennicutt98b} established the global Schmidt law in galaxies using 
the correlation between IR and CO;  \cite{rownd99} 
found a strong correlation between the H$_{\alpha}$ surface 
brightnesses and the CO integrated 
intensity; and, finally, a tight correlation between IR and HCN luminosities 
has been observed by \cite{gao04}.

The tightest correlation observed in spiral galaxies is that between the non-thermal 
radio and the FIR luminosity \citep{condon92}.
This correlation extends over five decades and holds for a remarkably wide variety 
of galaxy types \citep[e.g.,][]{yun01}. 
It applies not only globally, but also on kpc scales 
within the discs of individual galaxies (e.g., \citeauthor{boulanger88} \citeyear{boulanger88};
\citeauthor{bicay90} \citeyear{bicay90}; \citeauthor{murphy06} \citeyear{murphy06}; \citeauthor{tabatabaei07} \citeyear{tabatabaei07}). 
It has been established also that the CO emission in spiral
galaxies is well correlated
with the radio-continuum (RC) emission
on global and intermediate scales (e.g., \citeauthor{rickard77} \citeyear{rickard77};
\citeauthor{israel84} \citeyear{israel84}; \citeauthor{adler91} \citeyear{adler91}; 
\citeauthor{murgia02} \citeyear{murgia02}).
Recently, \cite{matteo05} and \cite{io06} extended the study
of the RC-CO correlation down to angular scale of $\sim 6''$ for a
sample of 22 spiral galaxies
selected from the Berkeley-Illinois-Maryland Association 3mm CO Survey 
of Nearby Galaxies (BIMA SONG; \citeauthor{regan01} \citeyear{regan01}; 
\citeauthor{helfer03} \citeyear{helfer03}). 
They found that
the RC-CO correlation is as tight as the global one,
down to sub-kpc size scales.

\begin{figure}[t]
\begin{center}
\includegraphics[width=8 cm]{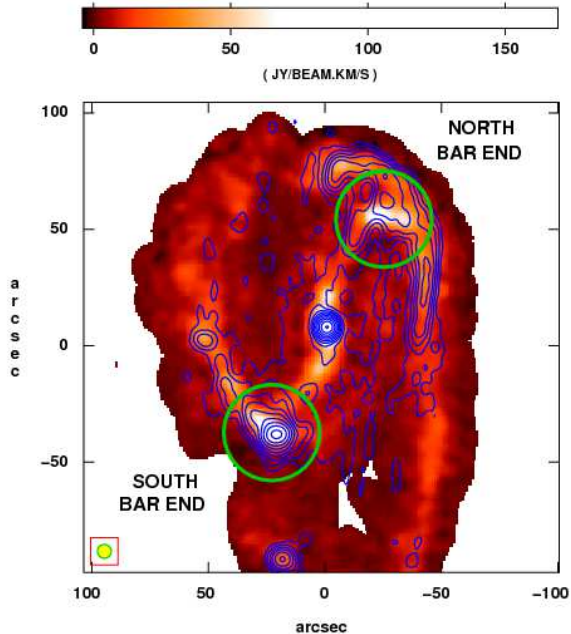}
\end{center}
\caption[]{Overlay of the VLA 1.4 GHz (contours) and BIMA CO (grey-scale) intensity.
The synthesized beam of both observations (6$''$) is shown in lower left corner.
The circles mark the north and south bar end regions we observed with the 
Plateau de Bure Interferometer and with VLA; their diameter represents the primary 
beam of a single PdBI antenna at 115 GHz.}
\label{point}
\end{figure}

\begin{figure*}[ht!p]
\begin{center}
\includegraphics[angle=270, width=0.75\textwidth]{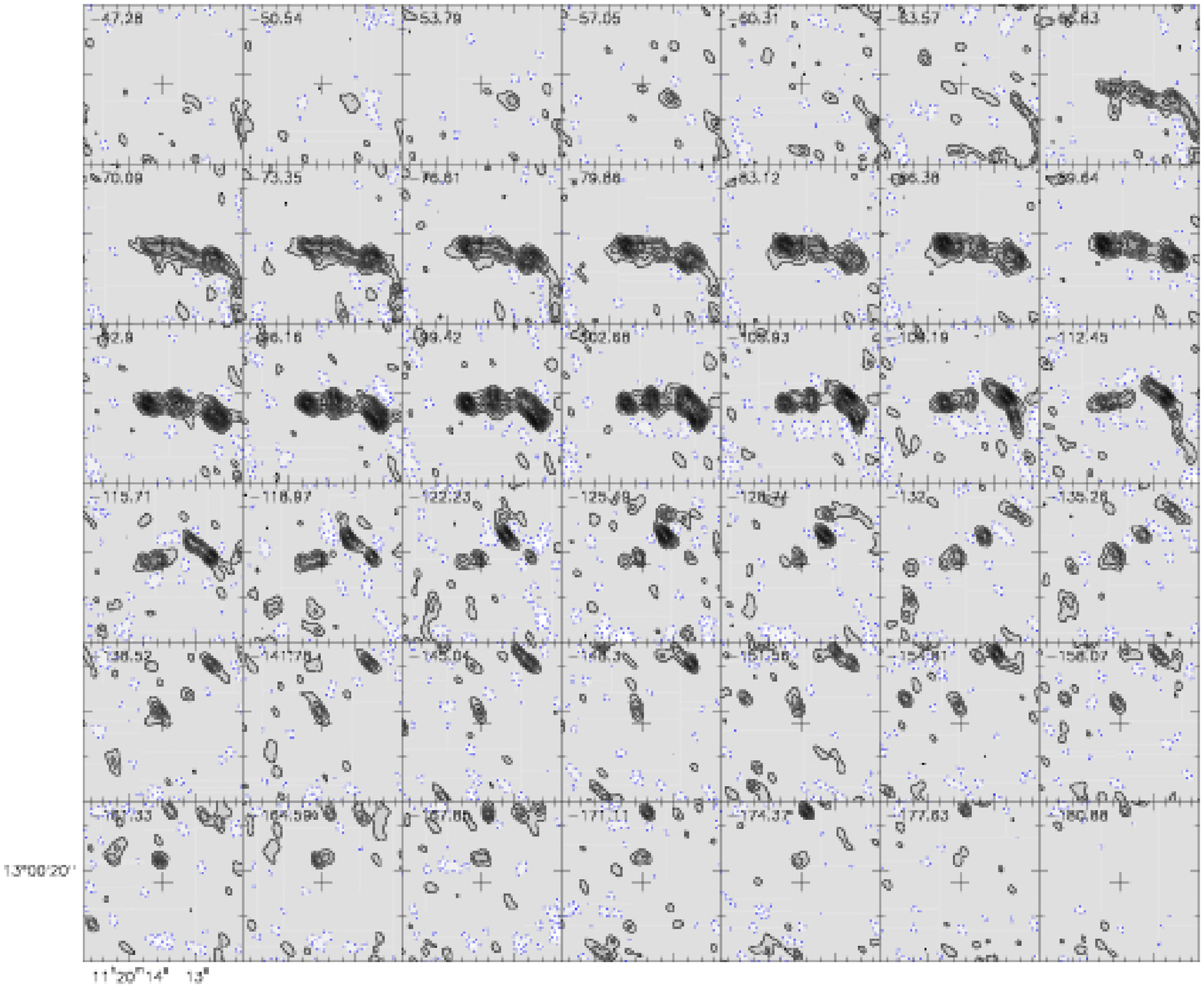}
\end{center}
\begin{center}
\includegraphics[height=8 cm, bb= 1 490 596 795]{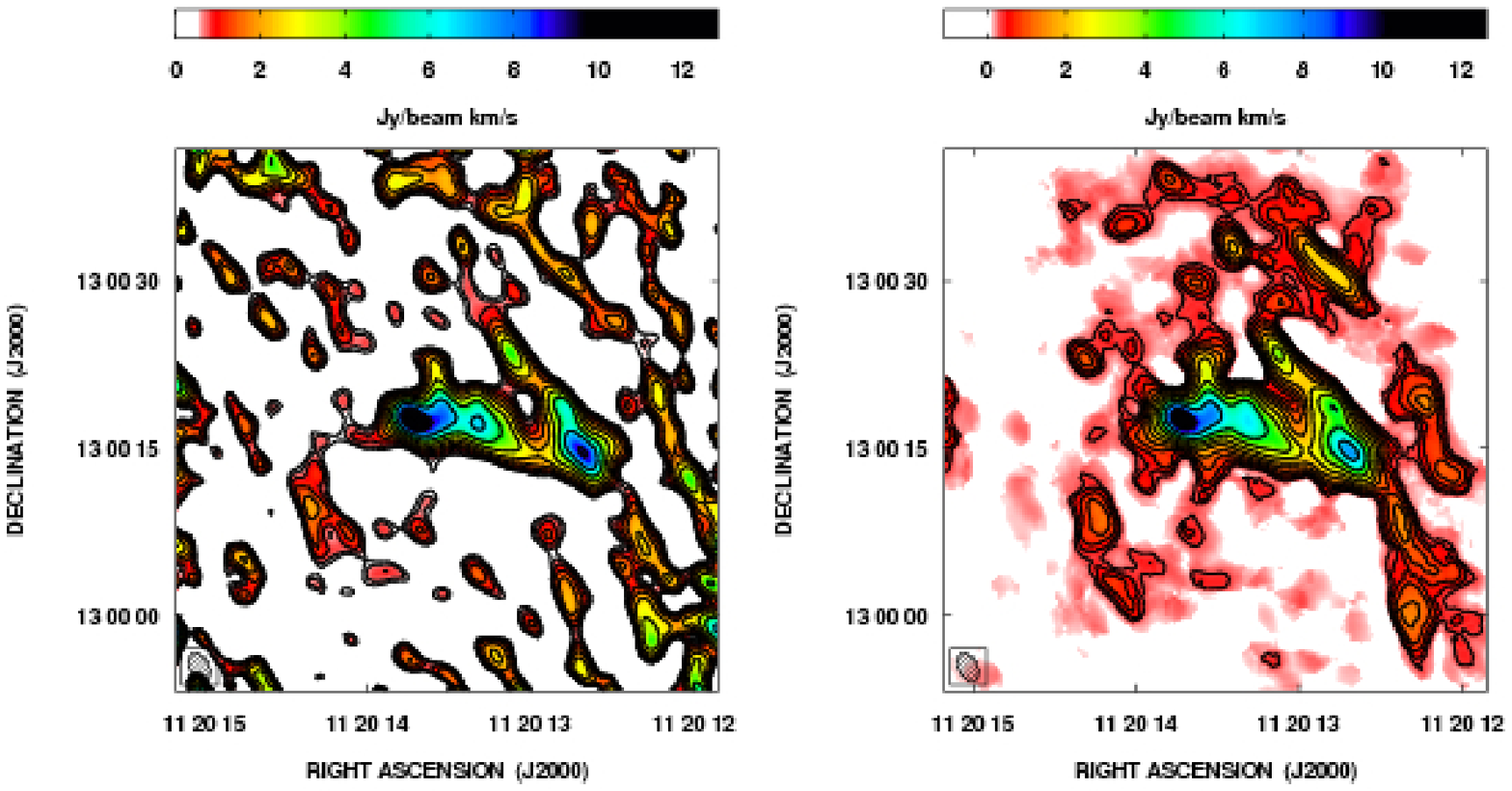}
\end{center}

\caption[]{{\bf North bar end of \object{NGC\,3627}}. {\it Upper panel}: $^{12}$CO(1-0) velocity-channel maps observed with the PdBI. 
The angular resolution is of  
2\farcs72$\times$1\farcs75 at PA=31\deg. 
We map an area of view of 35\arcsec (the diameter of 
the primary beam at 115 GHz is of 43\arcsec). The phase 
tracking center is indicated by a cross 
at $\alpha_{J2000}$=11$^h$20$^m$13.5$^s$ and 
  $\delta_{J2000}$=13\deg 00\arcmin 17\farcs7. Velocity-channels are 
displayed from v=$-$47.28 km s$^{-1}$ to v=$-$180.88 km s$^{-1}$ in steps of 
3.26  km s$^{-1}$. Velocities are referred to the LSR and the zero velocity corresponds to  v$_0$=712.6
km s$^{-1}$. Contour levels start from $-$30 mJy beam$^{-1}$ in step of 
30 mJy beam$^{-1}$. The rms noise is 6 mJy beam$^{-1}$ 
and only regions whose brightness is larger than   
5-$\sigma$ are shown.

{\it Lower panel}: North bar end of \object{NGC\,3627}: $^{12}$CO(1-0) integrated intensity.
The left panel shows the velocity integrated image.
Contour levels start from 0.4 Jy km s$^{-1}$ beam$^{-1}$ (the 1-$\sigma$ level is 0.13
Jy km s$^{-1}$ beam$^{-1}$) and scale 
by a factor of $\sqrt{2}$. The right panel shows the image obtained using the 
smooth-and-mask technique (see Sect.\ref{CO}). 
The grey-scale image shows emissions larger than the pixel 3 $\sigma$ level. 
The contours start from 0.3 Jy km s$^{-1}$ beam$^{-1}$, an 
average value of the rms in fainter regions, and scale 
by a factor of $\sqrt{2}$. }
\label{CO_nord}
\end{figure*}

\begin{figure*}[htp!]
\begin{center}
\includegraphics[height=15cm, width=0.65\textwidth,  angle=270]{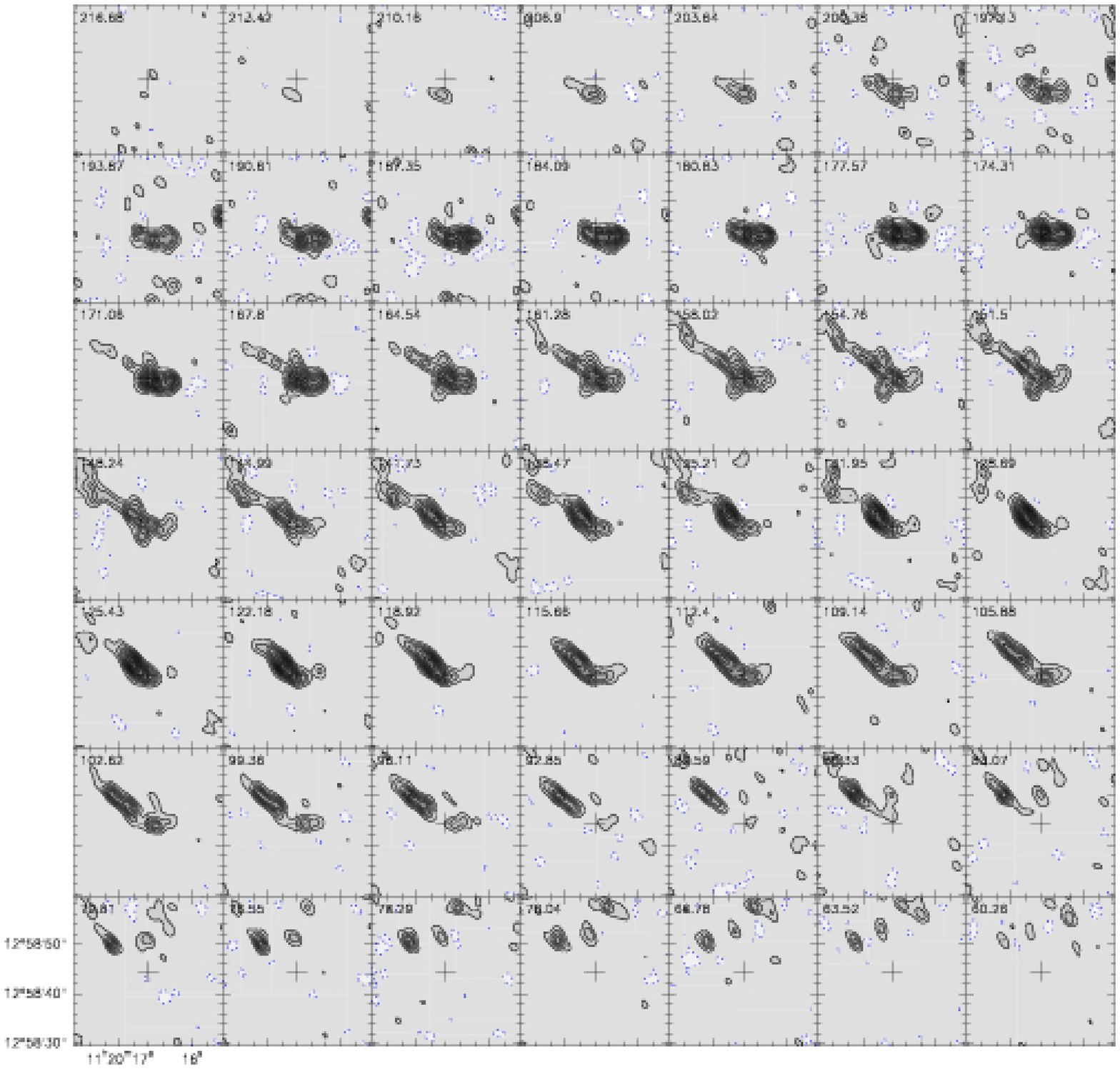}
\end{center}
\hfill
\begin{center}
\includegraphics[height=8 cm, bb= 0 490 596 795]{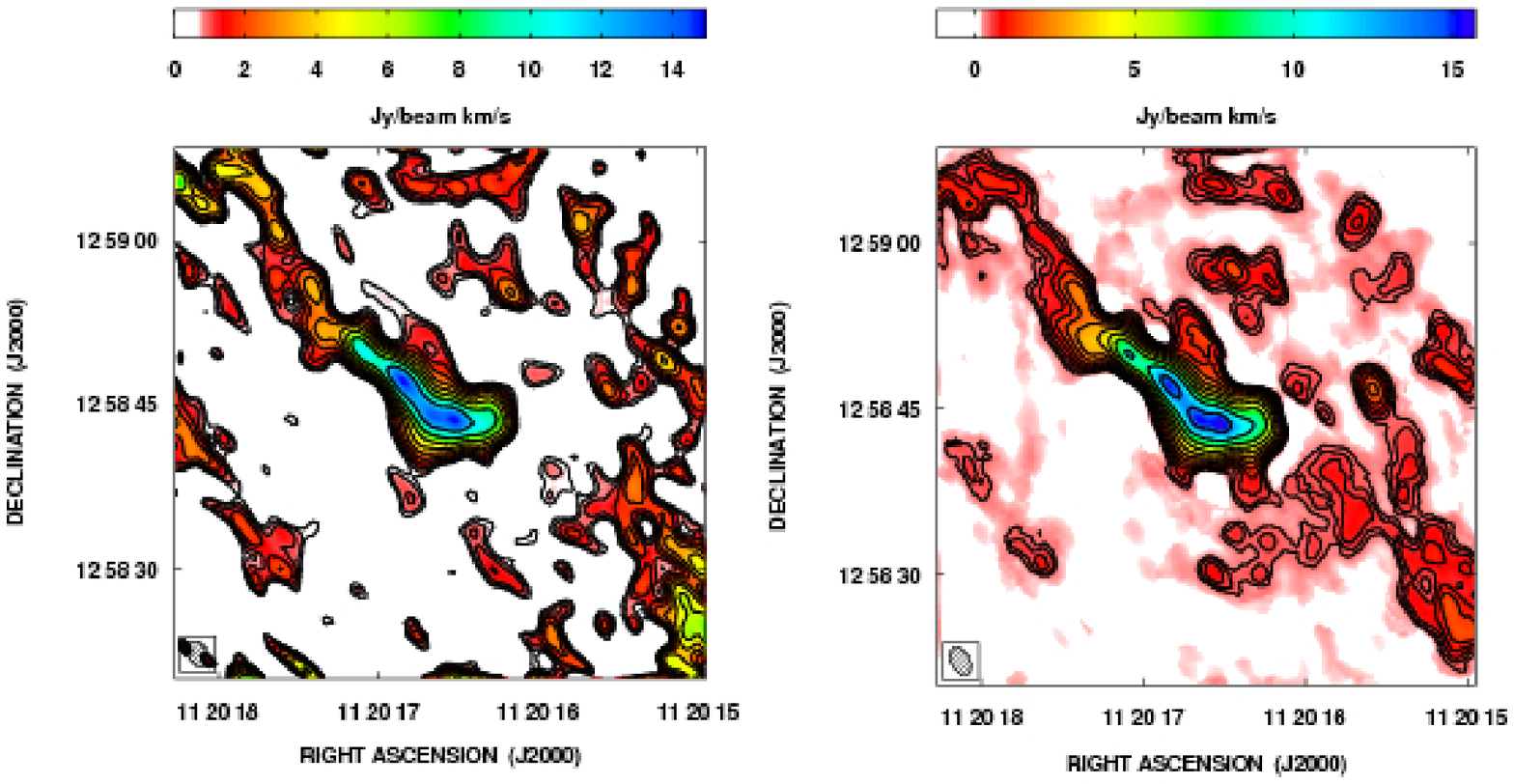}
\end{center}

\caption[]{{\bf South bar end of \object{NGC\,3627}}. {\it Upper panel}: $^{12}$CO(1-0) velocity-channel 
maps observed with the PdBI. The angular resolution of  
2\farcs75$\times$1\farcs73 at PA=31\deg. 
We map an area of view of 25\arcsec, i.e., about half of the diameter of 
the primary beam at 115 GHz. The phase tracking center is indicated by a cross 
at $\alpha_{J2000}$=11$^h$20$^m$16.6$^s$ and 
  $\delta_{J2000}$=12\deg 58\arcmin 44\farcs5. Velocity-channels are 
displayed from v=216.68 km s$^{-1}$ to v=60.26 km s$^{-1}$ in steps of 
3.26  km s$^{-1}$. Velocities are referred to the LSR and the zero velocity corresponds to v$_0$=712.6 
km s$^{-1}$. Contour levels start from -30 mJy beam$^{-1}$ with step of 
30 mJy beam$^{-1}$.The rms noise is 7 mJy beam$^{-1}$ 
and only regions whose brightness is 
above 4-$\sigma$ levels are shown.

{\it Lower panel}: $^{12}$CO(1-0) integrated intensity.
The left panel shows the velocity integrated image.
Contour levels start from 0.5 Jy km s$^{-1}$ beam$^{-1}$ 
(the 1-$\sigma$ level is 0.16 Jy km s$^{-1}$ 
beam$^{-1}$) and scale 
by a factor of $\sqrt{2}$. The right panel shows the image obtained using 
the smooth-and-mask technique (see Sect. \ref{CO}). The grey-scale image shows 
emission larger than the pixel 3 $\sigma$ level. 
 The contours start from 0.3 Jy km s$^{-1}$ beam$^{-1}$, an 
average value of the rms in fainter regions, and scale 
by a factor of $\sqrt{2}$.
 }
\label{CO_sud}
\end{figure*}

Despite the fact that the aforementioned local RC and CO studies on galaxies 
enhance differences between the slopes of the correlation 
from galaxy to galaxy, and inside the same galaxy, from region to region, 
they clearly indicate that the RC brightness is spatially correlated 
to the CO emission. 
So far, the scarce angular resolution of the FIR observations prevents 
detailed investigations of the FIR-RC correlation on sub-arcminute scales.
Nonetheless, recent high resolution {\it Spitzer} observations 
allow us the comparison of RC and CO emissions to shorter wavelengths 
than the FIR ones. 
In fact, using {\it Spitzer} 24 $\mu$m images, we found  
that RC is still correlated with this emission \citep{io06}.   
 The question  that we are trying to address is, what are the physical 
basis of these spatial correlations. 
The non-thermal RC is, for the most part, synchrotron emission 
that arises from the interaction of relativistic electrons with the ambient 
magnetic field in which they diffuse. These electrons are supposed to be 
accelerated in supernova remnants, which represent the ultimate phase of the 
life of massive stars. The reason why RC emission is so 
closely related to IR and CO is not so straightforward.
We know that the relativistic electrons diffuse from their 
birthplaces, therefore one would expect that the spatial correlations 
break down below the characteristic diffusion scale-length 
of the radiating electrons. A typical value of relativistic electrons 
diffusion scale in spiral galaxies is of the order of a few kpc 
(see also Sect.\,\ref{disc}).
Although, the value of the diffusion scale is extremely uncertain, 
depending on the propagation of cosmic-ray electrons along magnetic field lines, 
which is not fully understood. 

The recent result of a tight RC-CO-24 $\mu m$ correlations down to 
sub-kpc scale \citep{io06}
indicates either that we have not yet probed the spatial scale at which 
the correlations break down or that there is a mechanism of regulation 
that compensates the electrons diffusion.
With the aim of testing the former possibility 
we present the results obtained on one galaxy of the studied sample \citep{io06},
 \object{NGC\,3627}, observed at resolution of $\sim$  100 pc, by extending 
the study of the correlations between RC and CO emission to higher 
linear resolutions. 

We present new RC and CO observations, at a resolution of 2\arcsec, 
of two regions of the spiral galaxy \object{NGC\,3627}, 
obtained with the Very Large Array\footnote{The National Radio Astronomy
Observatory is a facility of the National Science Foundation operated
under cooperative agreement by Associated Universities, Inc.} (VLA) and 
the Plateau de Bure Interferometer (PdBI), respectively.  
The galaxy \object{NGC\,3627} is a promising target because it is a relatively nearby,
face-on galaxy, at the distance of 11.1 Mpc \citep{saha99}. At this distance 
1\arcsec\, corresponds to 53 pc.
Both RC and CO emissions are localized in a narrow bar 
structure of $\sim$300 pc in width. The emissions show a peak 
at the position of the nucleus, which 
extend along the leading edges of the bar forming two broad peaks at the 
bar ends, and then trail-off into the spiral arms.
\cite{io06} found that 24 $\mu$m-RC-CO correlations in this galaxy 
persist down to a spatial scales of 350 pc.
The new RC and CO observations are focused only on two regions of 
the bar far away from the nucleus of the galaxy (see Fig. \ref{point}).
As \object{NGC\,3627} is a Seyfert galaxy, we prefer to avoid the core whose RC 
emission is  likely contaminated  by the active galactic nuclei.

We report on the outcomes of new RC and CO 
observations of two regions of \object{NGC\,3627}, 
at resolution of $\sim$2\arcsec, corresponding to $\sim$ 100 pc 
at a distance of 11 Mpc. The resolution achieved is comparable to 
that of mid-IR {\it Spitzer} observations.
In order to add information about the connection between the RC emission
and other thermal emissions on spatial scale of 100 pc, we    
compare our new observations to archival 8, 5.8, 4.5 $\mu$m {\it Spitzer} and 
H$_{\alpha}$ observations.
This work is organized as follows:
in Sect. 2, we summarize new and archival observations,
 and give details of data reduction procedures.  
Data analysis and results are presented in 
Sect. 3. The results are discussed in Sect. 4, and a summary is given 
in Sect. 5.

\section{Observations and Data Reduction}

\subsection{CO data}
\label{CO}

The observations of the CO emission from \object{NGC\,3627} were carried 
out with the IRAM interferometer 
on March 20,\,2005, using the 6Cp  configuration of the array.
We observed simultaneously the J=1-0 and J=2-1 lines of $^{12}$CO 
at the bar ends of the galaxy (for details see Fig. \ref{point} and Tab. \ref{pos}); 
the beam size is 2\farcs7 $\times$ 1\farcs7 and 
1\farcs7 $\times$ 0\farcs8 
in the CO(1-0) and CO(2-1) line, respectively.

The spectral correlator was split in two IF centered at 114.997 and 
229.99 GHz, respectively. These are the transition rest frequencies of 
the two  $^{12}$CO lines for an assumed 
recession velocity of v$_0$(LSR)=712.6 km s$^{-1}$, which is 
the average of the 
recession velocities at the galaxy bar ends (Tab. \ref{pos}).
Using four partly overlapping 160 MHz-wide units, 
the correlator was configured to cover a bandwidth of 
580 MHz for each line, corresponding  
to 1512 km s$^{-1}$ and 756 km s$^{-1}$ at 115 GHz and 230 GHz, respectively.
We obtained visibilities with on-source integration times of 10 minutes 
framed by short ($\sim$ 2 min) integrations on the nearby 
quasars used as amplitude (3C84, 3C273) and phase (0923+392, 1055+018) calibrators.
Total observing times are reported in table \ref{time}.

We are primarily interested in  the CO(1-0) transition,
and combine the 
PdBI data with the available BIMA observations.
The $^{12}$CO(2-1) observations are heavily affected by the "missing flux" 
problem (see Sect. \ref{miss}), 
which makes them not reliable for a more quantitative analysis.
However, for the sake of completeness,  the CO(2-1) velocity channel maps 
and integrated intensity images 
are also reported in Appendix A. 

In Figs. \ref{CO_nord} and \ref{CO_sud} (upper panels), 
we show the velocity channel maps 
observed with PdBI in the north and south bar ends, correspondingly.
The rms derived from emission-free 
channels of 3.26 km s$^{-1}$ is of 6 mJy beam$^{-1}$ in the north bar end  
and of 7 mJy beam$^{-1}$ in the south bar end. No continuum emission was detected, 
down to 1 $\sigma$  rms noise level of 0.4 mJy beam$^{-1}$  (0.5  mJy beam$^{-1}$), 
for the north (south) bar end.
\begin{figure}[hb!]
\begin{center}
\includegraphics[height=8 cm]{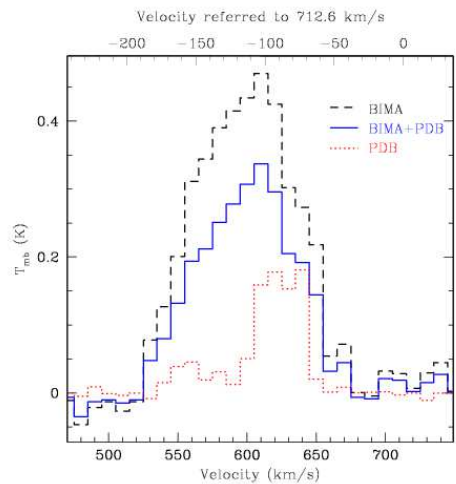}
\end{center}
\caption[]{Comparison of BIMA and PdBI spectra measured in 
the northern emission region. Data have been averaged  in RA-DEC plane, obtaining 
the profile along the velocity axis. 
The PdBI spectrum is shown in (red) dotted line, the BIMA in  
(black) short dashed  line and 
the short-spacing corrected spectrum in (blue) solid line.
}
\label{imm_N}
\end{figure}

\begin{figure*}[t]
\begin{center}
\includegraphics[height=9 cm, bb=0 510 595 800]{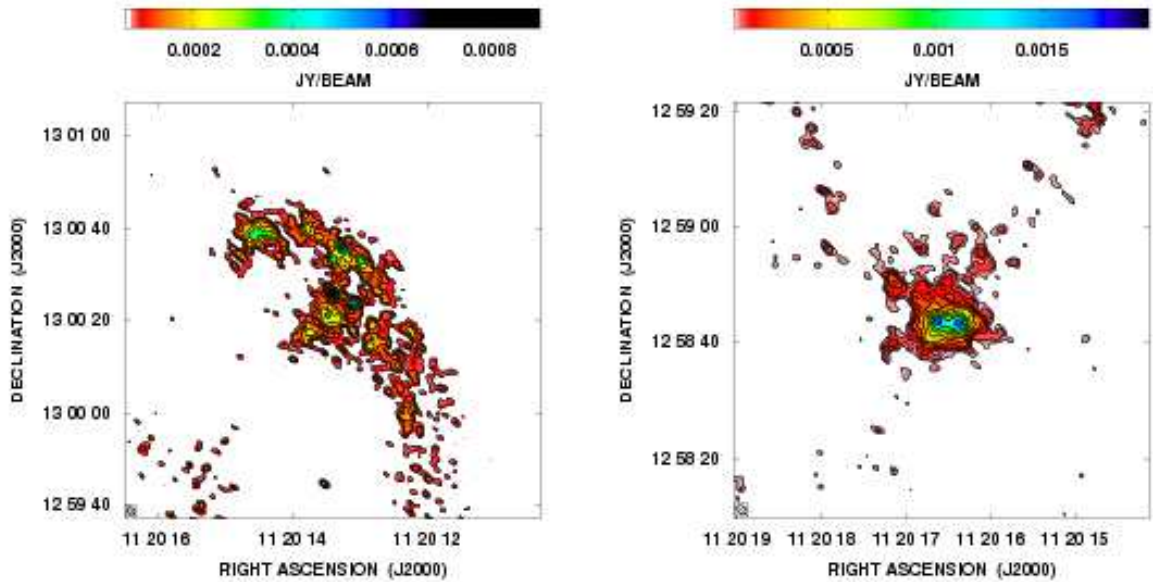}
\end{center}
\caption[]{ 
VLA-A array images of north bar end ({\it Left}) and south bar end ({\it Right}) 
of \object{NGC\,3627}. The angular resolution is of 1\,\farcs9 $\times$ 1\,\farcs3 at PA=49\degr.
Contours level start from 72 $\mu$Jy beam$^{-1}$ and 66  $\mu$Jy beam$^{-1}$ (3$\sigma$), 
respectively, and scale by a factor of $\sqrt{2}$. 
}
\label{VLA_A}
\end{figure*}

In Figs. \ref{CO_nord} and \ref{CO_sud} (lower panels), we present 
the  $^{12}$CO integrated intensity images for the two 
observed regions. The left panel shows the integrated intensity obtained 
by summing all channels in the data cube. The 1-$\sigma$ level is given by the 
noise estimated in the free channel multiplied by $\sqrt{N} \Delta v$, 
where N is the number 
of channels summed and the $\Delta v$ is the width of each channel (3.26 km s$^{-1}$): 
in the north and south regions, the 1-$\sigma$ levels are 
0.13 Jy beam$^{-1}$ km s$^{-1}$ and 0.16 Jy beam$^{-1}$ km s$^{-1}$, respectively.
The right panel shows the integrated intensity formed using the 
smooth-and-masked technique \cite[see][]{helfer03}, which is 
very effective in showing low level emission in the map.
First, we smoothed the data cube by a Gaussian 
with FWHM=6\arcsec\, and we created a mask accepting all pixels in each 
channel map where the signal 
 was stronger than 3 $\sigma$ rms noise level of the smoothed channel.
Second, we selected in the full-resolution data cube only those pixels that
 with non-zero values in the mask file and summed them to obtain the integrated 
intensity image. Finally, we multiplied the pixel values 
by the velocity width of an individual channel 
(3.26  km s$^{-1}$) to generate the image in the units of 
Jy beam$^{-1}$ km s$^{-1}$. 
This technique implies a variation of the pixel-by-pixel rms 
in the CO integrated intensity 
image. The plots in Figs. \ref{CO_nord} and \ref{CO_sud} show 
contours starting from 0.3 Jy beam$^{-1}$ km s$^{-1}$, the average value of the 
rms in fainter regions, whereas the grey-scale images show emissions above 3 times the 
rms measured pixel-by-pixel.

\begin{table}
\begin{tabular}{lccc}
\hline \hline
Source    & RA & DEC & LSR Velocity \\
& (J2000.0)   & (J2000.0)     & (km/s)\\
\hline
NGC3627{\bf{N}} & 11:20:13.5 &  +13:00:17.7 & $+$600\\
NGC3627{\bf{S}} & 11:20:16.6 &  +12:58:44.5 & $+$825\\
\hline
\end{tabular}
\caption{Pointing positions and local standard of rest velocities for the two 
bar ends.}
\label{pos}
\begin{tabular}{lcccc}
\hline
\hline
    & VLA$^*$ & PdBI$^*$ & $\sigma_{1.4}$& $\sigma_{\rm CO(1-0)}$  \\
&hr  &hr & $\mu$Jy bm$^{-1}$ &Jy bm$^{-1}$ km s$^{-1}$ \\
\hline
NGC3627{\bf{N}} & 6 & 8 &22  &0.13 \\
NGC3627{\bf{S}} & 6 & 8 &24 &0.16\\
\hline
\end{tabular}
\caption{Observations summary. Observing times and rms levels for 
VLA and PdBI observations (see Sect.\ref{CO} and \ref{RC}).}

$^{*}$ Observing times include overheads.
\label{time}
\end{table}

\subsubsection {BIMA and PdBI data combination}
\label{miss}

Our PdBI observations are affected by the ``missing flux'' problem.
The baselines range from 16 to 330 meters.
The observed regions have extension of 
$\sim$ 25\arcsec, larger than half the primary beam size of PdBI 
antennae at 115 GHz (43\arcsec).
In this case,  the PdBI observations 
may have missed about 50\% of the flux \citep{helfer02}.

To reduce this problem we combined the PdBI data with those of the BIMA SONG. 
The BIMA interferometer has smaller antennas than PdBI (dish diameter is 6 m), 
and minimum dish separation of 8 m, the largest angular size detectable at 115 GHz 
being  $\sim$67\arcsec.  
In addition, data obtained with NRAO 12 m telescope have been incorporated 
to interferometric maps in BIMA SONG images, therefore these images are 
not missing any flux \citep{helfer03}.
The combination should be performed by merging   
PdBI and BIMA u-v data together and then combining this with the 12 m data for 
the deconvolution (see \cite{helfer03} for the application of this method to the 
BIMA SONG).
In our case, using this method is very difficult, 
since we have two discrete PdBI  
fields across a region containing extended emission 
observed in a multi-pointings mode with BIMA. 
We decided to combine PdBI and BIMA data in the image plane, using 
the technique  implemented in the 
MIRIAD (Multichannel Image Reconstruction Image Analysis and Display) 
task IMMERGE \citep{miriad95}. 
This technique requires caution, nevertheless, 
\cite{wobli00} show that this method agrees well with the 
``linear combination'' described in \cite{stanimirovic99} in which 
data are combined before deconvolution and then deconvolved with a composite beam.
The task IMMERGE applies the ``Fourier'' technique: the maps are transformed 
to the Fourier plane and merged together.  
In a specified annulus of the Fourier domain, where both images should agree, 
we calculated the flux calibration factor, which minimize differences 
between the data of the two resolutions.
Since the ranges of baseline length of the 
PdB and BIMA interferometer
are\, 20-230 \,m and \ 8-87 m, respectively, 
the annulus used is between 30 and 80 meters.
Since the primary beams of the two images are different, the low-resolution image is 
tapered to match any residual primary beam response in the 
high-resolution one.

For the combination of PdB and BIMA data, we regrid the PdBI cubes 
to the same velocity grid as 
the BIMA+12m cubes. 
For comparison between the PdBI and BIMA data, 
in Fig.  \ref{imm_N} we show the average 
spectrum measured in the northern region. Data have been averaged in 
RA-DEC plane, obtaining the profile along the velocity axis.  
In this plot it is evident that
the inclusion of the short-spacing data has a noticeable impact on the flux of 
channels whose velocities lie between $-$120 and $-$180 km s$^{-1}$, 
i.e., those with more diffuse emission.
Missing flux still remains an issue in our CO combined images. 
Nevertheless, we believe that the 
adopted combination technique is sufficient for our purpose 
of analyzing the moment-0 images. 
In particular, \citeauthor{helfer03} (\citeyear{helfer03}; their Fig. 53) 
indicate that the flux is uniformely lost by BIMA throughout the galaxy.
Hence, if a similar behavior can be expected for our combined maps, the missing flux 
would only produce a horizontal shift of the point-to-point RC-CO correlations 
shown below (see Figs. \ref{N_2arcsec} and \ref{S_2arcsec}), without affecting  
the slopes.  Such an effect would not invalidate the overall results.

\subsection{Radio continuum data}
\label{RC}

\begin{figure*}[ht!]
\begin{center}
\includegraphics[width=\textwidth, bb=12 492 568 682]{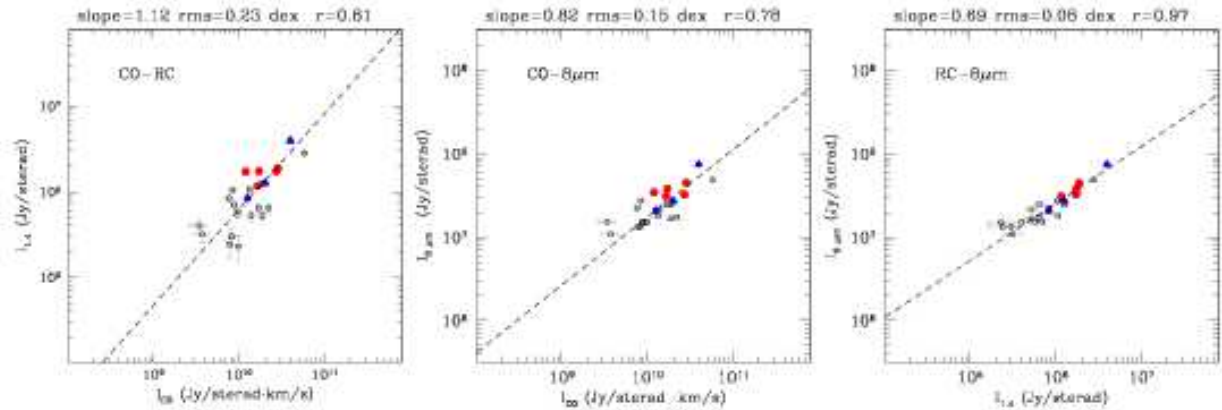}
\end{center}
\caption[]{Comparison between I$_{1.4}$, I$_{\rm CO}$
and 8$\mu$m mid-IR emissions. Brightnesses are measured by 
averaging all pixel values within circles of 
18\arcsec in diameter recovering the entire galaxy.  
Filled blue triangles represent the measures in 
the south bar end, filled red circles are those in the north bar end. 
}
\label{6arcsec}
\end{figure*}

We observed the 1.4 GHz emission from the two regions of \object{NGC\,3627} 
with the VLA in its A configuration on April, 11, 2006, spending 
$\sim$ 3.5 hours on each target .
We observed with a bandwidth of 50 MHz for the two IF channels centered at 1.465
and 1.385 GHz.

The data were calibrated and imaged with the NRAO package AIPS following
 standard procedures:  Fourier-Transform, Clean and Restore.
Self-calibration was applied to minimize the effects of phase variations.
 The flux densities were brought to the scale of \citet{baars77}  using the
calibrators 3C138 and 3C286.
The final cleaned maps were restored with beams of 1\,\farcs9$\times$1\,\farcs3.
Figure \ref{VLA_A} shows the VLA-A array maps obtained for the two observed  
regions, produced using natural weighting. The final noise of total intensity 
images obtained for the north and south bar end region is of 22 and 24 $\mu$Jy beam$^{-1}$, 
respectively.

To improve the sensitivity to large scale structures, the A-array data  were combined 
with D and B-array data in the visibility plane, using the AIPS 
task DBCON. The B-array observations are described in \cite{io06}, while   
those of the D-array are archival data.
The combined image has a resolution of 2\,\farcs07 $\times$ 1\,\farcs43 and a 
rms level of 21 $\mu$Jy beam$^{-1}$.

For the comparison with CO and IR images, we convolved the A+B+D-array VLA
 images  to the beam of both BIMA and PdBI observations, 
i.e., 7\,\farcs3 $\times$6\arcsec and 2\,\farcs7 $\times$1\,\farcs8,
 respectively. The derived low and high resolution images have a
 1-$\sigma$ noise of  136  $\mu$Jy beam$^{-1}$ 
and 27 $\mu$Jy beam$^{-1}$, correspondingly.

\subsection{Archival data}
\label{sect:arch}
The spiral galaxy \object{NGC\,3627} has been observed as part of the {\it Spitzer} Infrared 
Nearby Galaxies Survey \citep[SINGS][]{sings03},
and IR {\it Spitzer} images. In addition, H$_\alpha$ images 
have been released in the context of the SINGS Second Data Delivery, April 2005.

The observations obtained with the Infrared Array Camera \citep[IRAC;][]{fazio04}
 at 4.5, 5.8, and 8 $\mu$m  have resolutions of 
1\,\farcs2, 1\,\farcs5, and 1\,\farcs8, respectively.
The H$\alpha$ images, obtained  
at Kitt Peak National Observatory (KPNO) 2.1 m telescope, 
using a narrow-band filter centered at 6618 \AA,
have a PSF of 1\,\farcs9.
For details on {\it Spitzer} and H$\alpha$  observations and 
data processing see delivery document for SINGS 
Second Data Delivery\footnote{http:$//$data.spitzer.caltech.edu$/$popular$/$sings$/$20050506$\_$enhanced
$\_$v1$/$Documents$/$sings$\_$second$\_$delivery$\_$v2.pdf}.

The resolution of IRAC and H$\alpha$ images is only slightly smaller than that 
 of our RC and CO images, making the comparison between these sets  
of images consistent.
To this end, we convolved the IRAC and H$\alpha$ images
to the largest beam and  performed the 
geometric transformation of images to match with the pointings of 
RC and CO observations.

We compare the RC and CO emission with 
the dust emission at mid-infrared wavelengths.
The images at $\lambda\,24\,\mu$m can be considered pure 'dust' images, 
with negligible contributions from the photospheric emission of stars and 
from nebular emission, whereas  
the IRAC wavelenghts (3.6, 4.5, 5.8 and 8 $\mu$m)
 are dominated by stellar emission.
In particular, the 4.5 $\mu$m emission is dominated by photospheric 
emission from stars, but can also contain 
a component of hot dust emission. In 5.8 and 8 $\mu$m images, however, 
the emission from 
PAH becomes significant.
In order to obtain the distribution of dust at these wavelengths 
by subtracting the stellar contribution, we used the 
recipe of \cite{helou04}. These authors assume that all the 3.6 $\mu$m emission 
is due to stars and extrapolate this component to longer wavelengths 
using stellar population 
modeling. To reproduce the star contribution at  4.5, 5.8, and 8 $\mu$m,  
\cite{helou04} scale the 3.6 $\mu$m map by 0.596, 0.399, and 0.232, respectively, 
and subtract these scaled maps from the observed maps pixel by pixel to 
yield the "dust maps". 
Since this method may underestimate the dust component e.g., 
the 3.6 $\mu$m images can contain a hot dust component in addition to photospheric 
emission we judge this approach as a reliable one for our purposes. In fact, 
the hot dust emission has an impact below few percent on the dust-only 
images \citep{calzetti05}.

\section{Data analysis and results}
\label{an_res}

\begin{figure*}
\begin{center}
\includegraphics[height=8 cm, bb=24 540 569 825]{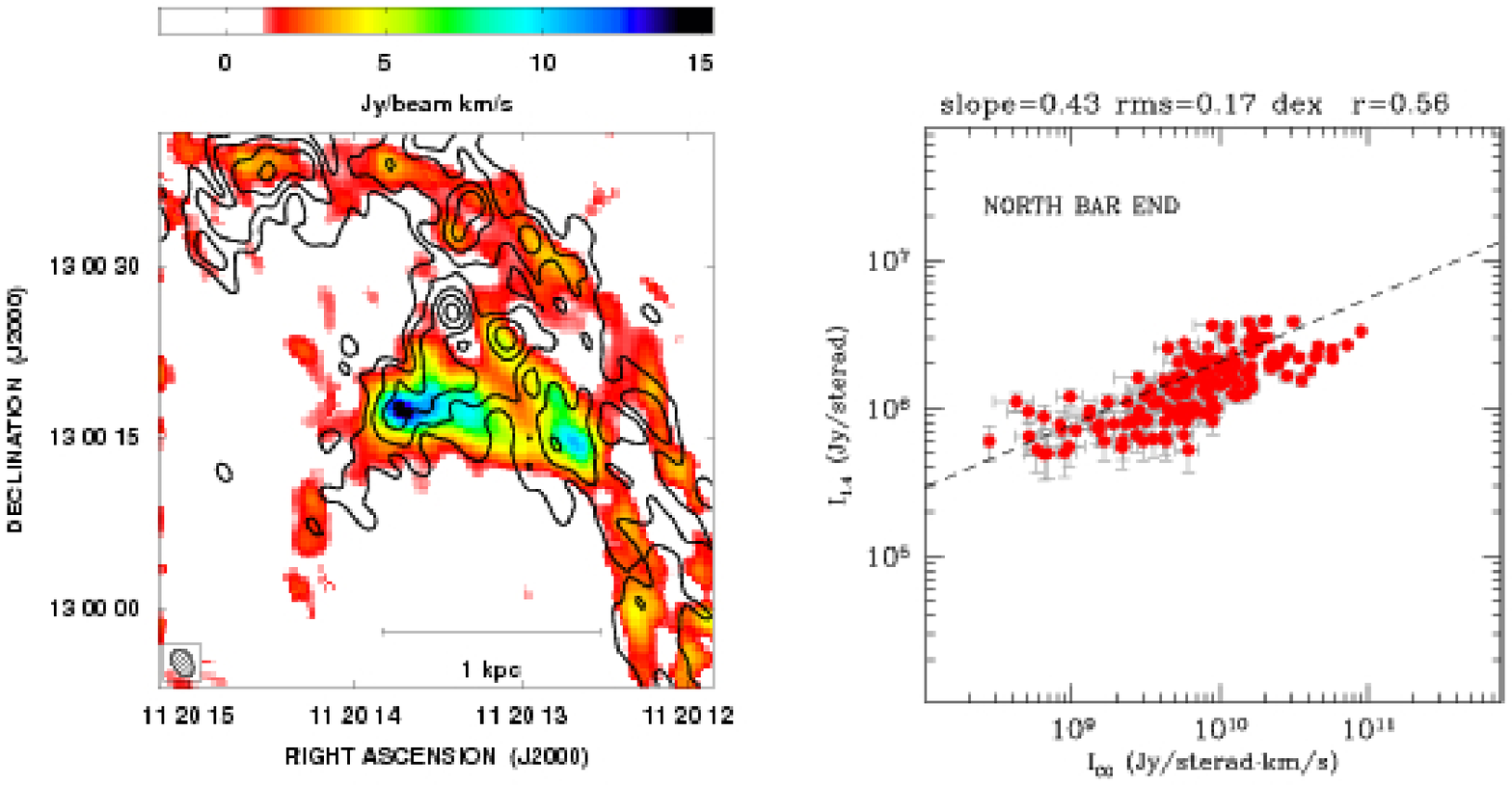}
\end{center}
~\
\caption[]{{\it Left}:
Overlay of RC brightness (contours) on  CO integrated intensity (grey-scale image) for the 
north bar region observed at a resolution of 2\,\farcs74$\times$1\,\farcs76.
Contours level start from 0.2 mJy beam$^{-1}$ (5$\sigma$)  and scale 
by a factor of $\sqrt{2}$. Note that the spatial coincidence between the two emissions is 
maintained well below the kpc scale.
 {\it Right}: I$_{1.4}$ versus I$_{\rm CO}$. Brightnesses are measured by 
averaging all pixel values within beam-sized boxes recovering the entire region of emission.
Only points above 3 $\sigma$ level have been plotted.
}
\label{N_2arcsec}

\begin{center}
\includegraphics[height=8 cm, bb=24 540 569 825]{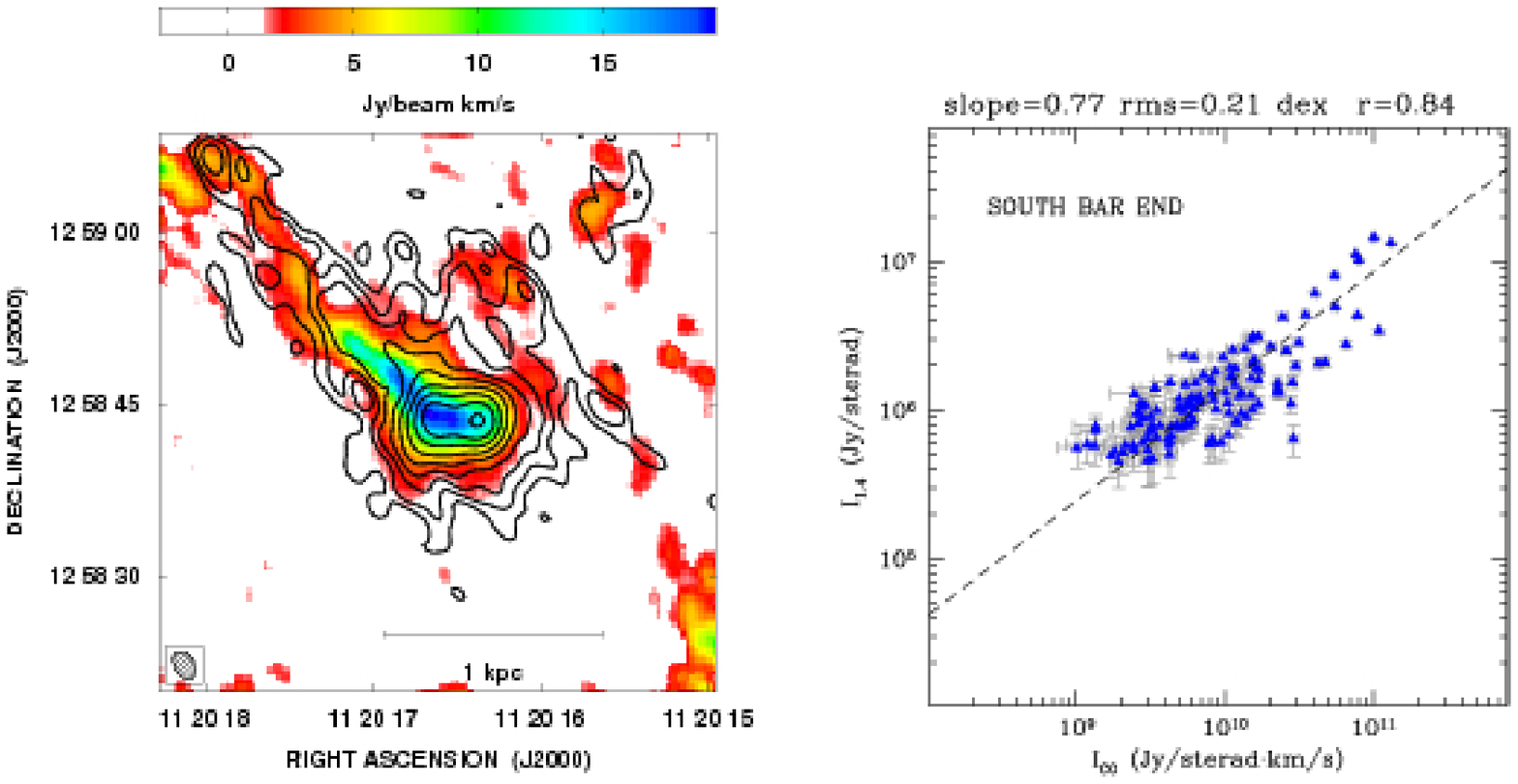}
\end{center}
~\
\caption[]{ {\it Left}:
Overlay of RC brightness (contours) on  CO integrated intensity (grey-scale image) for the 
south bar region observed at resolution of 2\,\farcs74$\times$1\,\farcs76.
Contours level start from 0.15 mJy beam$^{-1}$ (5$\sigma$)  and scale 
by a factor of $\sqrt{2}$. Note that the spatial coincidence between the two emissions is 
maintained well below the kpc scale.
 {\it Right}: I$_{1.4}$ versus I$_{\rm CO}$. Brightnesses are measured by 
averaging all pixel values within beam-sized boxes recovering entire region of emission.
Only points above 3 $\sigma$ level have been plotted.
}
\label{S_2arcsec}
\end{figure*}

\begin{figure}
\includegraphics[scale=0.6, bb=64 106 496 770]{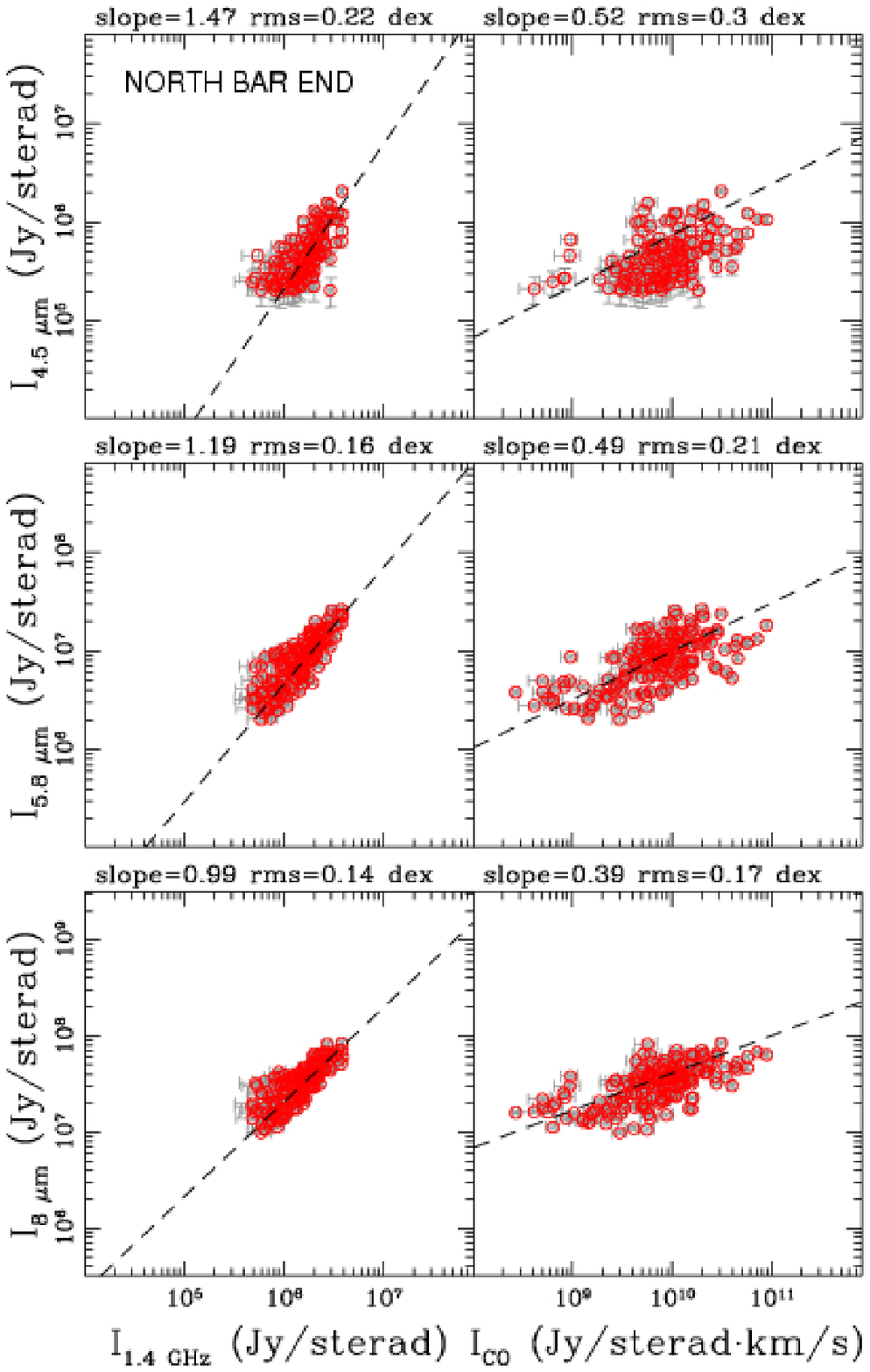}
\caption[]{Correlations between RC (left) and CO (right) brightness and 4.5\,$\mu$m , 5.8\,$\mu$m and  
8\,$\mu$m  brightness in the north bar end region.
}
\label{nord_8plot}
\end{figure}
\begin{figure}
\includegraphics[scale=0.6, bb=64 106 496 770]{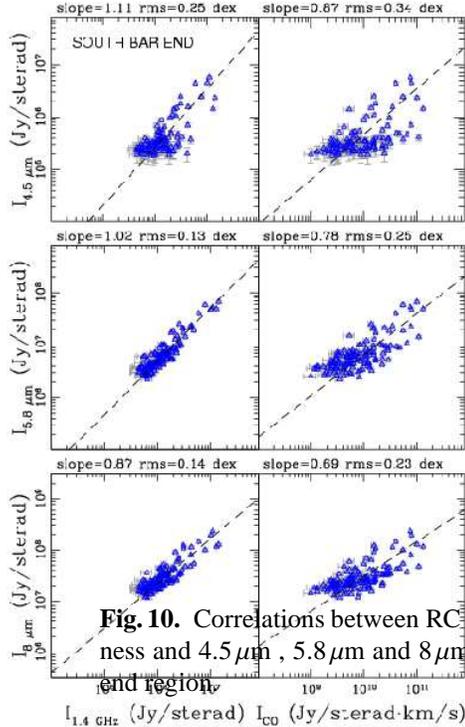}
\caption[]{ Correlations between RC (left) and CO (right) brightness and 4.5\,$\mu$m , 5.8\,$\mu$m and
 8\,$\mu$m 
brightness in the south bar end region. 
}
\label{sud_8plot}
\end{figure}

We extend the study of the RC-CO-24 $\mu$m correlations obtained for a spatial scale of 6\arcsec,
 reported in \citet{io06},
 to the 8\,$\mu$m {\it Spitzer} data, convolving 
8\,$\mu$m images to the BIMA resolution: 7\,\farcs3 $\times$ 6\arcsec.
To obtain a general view of the correlations across the entire galaxy, we overlaid a grid of  
circles with a diameter of 18\arcsec\, across the spiral structure. Then, we calculated 
the 1.4 GHz RC brightness ($\rm I_{1.4}$), the CO integrated intensity ($\rm I_{\rm CO}$) and the mid-IR 
brightness ($\rm I_{8 \mu m}$) 
by averaging all pixel values within each circle.
The correlations obtained are shown in 
Fig. \ref{6arcsec}. 
We fitted the observed data with a power law parametrized as $y(x)=c \cdot x^{slope}$,
taking into account the errors of both x and y.
The results of this fit and the Pearson coefficient of the 
correlations are reported as labels on the plots shown in Fig. \ref{6arcsec}.
The correlations of the 8\,$\mu$m emission with both RC and CO have scatters 0.08 and 0.15 dex, respectively.
This values are even 
 lower than that of  RC-CO correlation. In particular the RC-8\,$\mu$m correlation 
has a Pearson coefficient near 1.
The slopes of the correlations are very different, depending  on 
which variable is associated to the x and y-axis. To provide a more complete description of 
the RC-CO-8\,$\mu$m  
correlation, we introduce a three-dimensional view of it in Sect. \ref{3D}.

In Figs. \ref{6arcsec}, 
points corresponding to northern and southern regions,
 observed at high resolution with the PdBI and the VLA in this work, are 
marked with  filled blue triangles and filled red 
circles, respectively. 
In the following section, we report on 
the correlations studied at 2\arcsec\, resolution in these two regions.

\subsection{The mid IR-RC-CO point-to-point correlation at 2\arcsec}

In order to compare point-to-point  RC, CO, and mid-IR brightnesses in the 
observed regions, we overlaid on the images regular grids of rectangular boxes, whose 
dimensions coincide with the beam of observations (corresponding to a spatial scale 
of 100 pc), and we  averaged all pixel values within each box.
We performed the power law fitting described above, determining 
for each correlation the corresponding slope and normalization parameter.
The results of this fit are reported as labels on the plots.

Figures \ref{N_2arcsec} and \ref{S_2arcsec} show the  overlay of RC brightness 
contours on  CO integrated intensity  images and  
the plot of $\rm I_{1.4}$ versus $\rm I_{\rm CO}$ for both 
regions.
The overlay of the emissions shows that CO and RC emissions 
distribute very similarly in both the north and south bar ends,
 despite the presence of minor local differences. 
In particular, both RC and CO emissions follow the same 
filamentary distribution in both regions.
The plot of the point-to-point correlations 
 show that in the north bar end the RC emission spans only one order of magnitude,   
 whereas the CO emission varies over two orders of magnitude. 
 In the south bar end instead, both RC and CO emissions  vary  
by a comparable factor. This results in a flatter slope in the northern region (0.43) 
than in the southern one (0.77). The Pearson's correlation coefficient, r, is also 
different in the two regions (r=0.56
in the north bar end and r=0.84 in the south bar end), indicating that the RC is less 
correlated to the CO in the northern region than in the southern one. 
The scatter of both correlations (0.17 dex and 0.21 dex in the northern and southern region, respectively)
 is similar to that observed 
 in the 6\arcsec\, correlations presented in  \cite{io06}. 

\begin{figure*}
\begin{center}
\vspace{0.2cm}

\includegraphics[width=0.75\textwidth, bb=50 632 499 832]{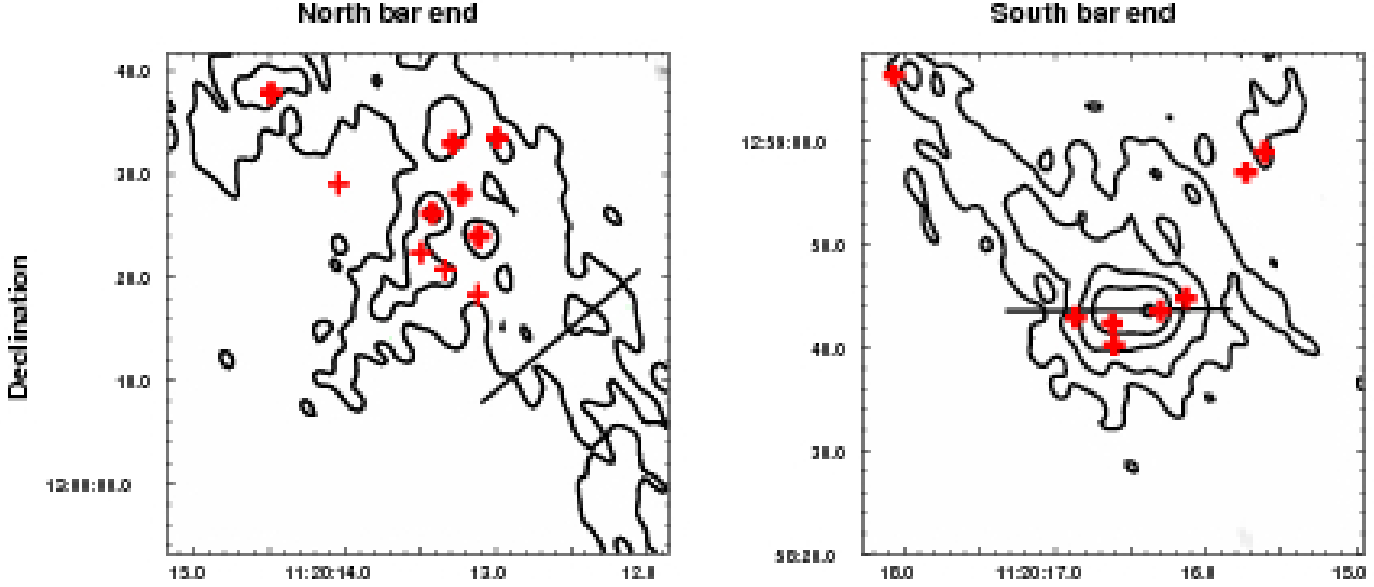}
\includegraphics[width=0.75\textwidth, bb=50 625 499 832]{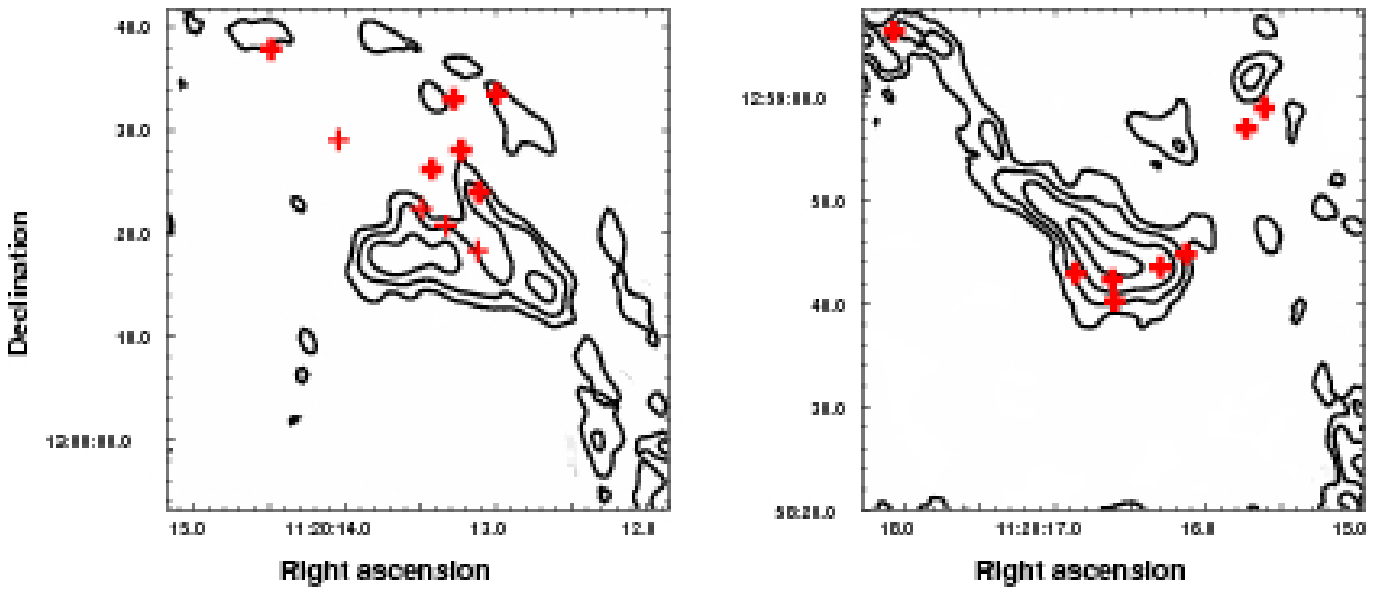}

\end{center}
\caption[]{Comparison between radio continuum (upper panels) and CO 
(lower panels) maps with 
HII regions positions. RC and CO contours start from 0.2 mJy beam$^{-1}$ 
and 2 Jy beam$^{-1}$
 km s$^{-1}$, respectively, and scale by a factor of 2, to enhance point sources.
Red crosses represent positions of brightest HII regions in the 
north (left panels) end, south (right panel) bar end. For a complete list of HII 
regions in \object{NGC\,3627},  
see \cite{hodge74}. Black lines in upper panels show the positions of two slices  for 
the comparison of emission profiles (see Sect. 4).
 }
\label{optha}
\end{figure*}

We performed the same point-to-point analysis using the 4.5, 5.8, and 8 \,$\mu$m 
``dust maps'' 
(see Sect. \ref{sect:arch}), 
convolved to the same beam of RC and CO images.
The results for north and south bar ends are reported in Figs. \ref{nord_8plot}
 and \ref{sud_8plot}, respectively.
In the two regions, it appears that both the RC and mid-IR emissions vary by about one 
order of magnitude, whereas the 
CO varies over almost three orders of magnitude, producing flat CO-mid IR correlations.
The RC is well correlated with dust emissions of both 5.8 and 8 \,$\mu$m: the resulting correlations
are nearly linear and scatters vary from 0.16 to 0.13 dex. 
The RC-4.5 \,$\mu$m correlation 
has a higher scatter, in particular in the southern region.
We report, in Table \ref{tab:corr2}, the fitted slopes, rms of 
the 2\arcsec\,correlations,  
and the Pearson's coefficients. 

\begin{table} 
\begin{tabular}{llcccccccc} \hline \hline
\multicolumn{2}{c}{}&&
\multicolumn{3}{c}{\bf North bar end}& &
\multicolumn{3}{c}{\bf South bar end}\\ \hline
 x&y&& slope&rms&r && slope& rms & r \\
\hline
\rule{0.0em}{1.3em}I$_{\rm CO}$&I$_{\rm RC}$&& 0.43 &0.17&0.56 &&0.77&0.21&0.84 \\
I$_{\rm RC}$&I$_{4.5 \mu m}$&& 1.47&0.22&0.69 && 1.11&0.25&0.74\\
I$_{\rm RC}$&I$_{5.8 \mu m}$&& 1.19&0.16&0.85 && 1.02&0.13&0.95\\
I$_{\rm RC}$&I$_{8 \mu m}  $&&  0.99&0.14&0.83 &&0.87&0.14&0.84\\
I$_{\rm CO}$&I$_{4.5 \mu m}$&& 0.52&0.3&0.41 &&0.87&0.34&0.53\\
I$_{\rm CO}$&I$_{5.8 \mu m}$&& 0.49&0.21&0.40 &&0.78&0.25&0.77\\
I$_{\rm CO}$&I$_{8 \mu m}  $&& 0.39&0.17&0.52 &&0.69&0.23&0.67\\[3pt]
\hline \hline
\end{tabular}
\caption{Coefficients of IR-RC-CO 2\arcsec correlations. Observed data have been fitted 
with the function $y(x)=c \cdot x^{slope}$. Slope is the result of the 
fit, rms is expressed in dex, and r is the Pearson's correlation coefficient.
}
\label{tab:corr2}
\end{table}

\begin{figure*}[htpb]
\begin{center}
\includegraphics[width=9cm, angle=270]{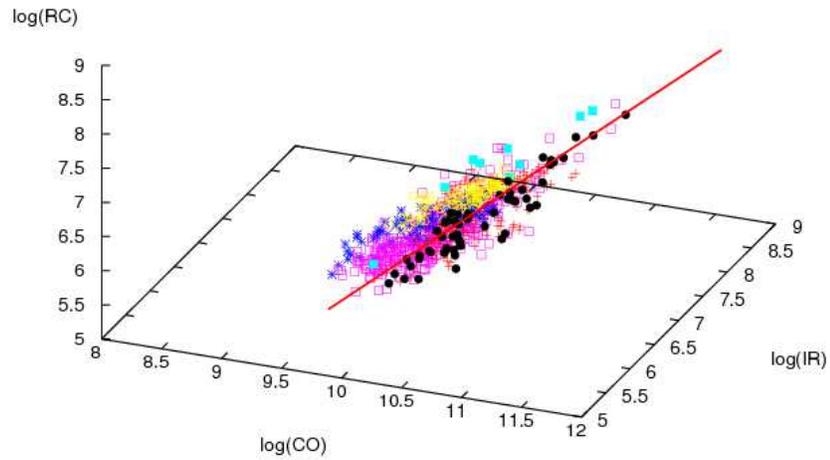}

\caption[]{{\bf RC, CO, and 24 \,$\mu$m 3D correlation.} The x, y, and z axis are the log(CO), log(IR) (24 $\mu$m),  
and log(RC), respectively. Data at 6\arcsec resolution of six galaxies 
reported in Table \ref{tab:tetafi} are 
shown and the line plotted is the result of the fit for all data. 
The scatter of this fit is 0.25 dex.
}
\label{3Dline_24m}
\end{center}
\end{figure*}

\begin{figure*}[htpb]
\includegraphics[width=\textwidth, bb=0 211 582 842]{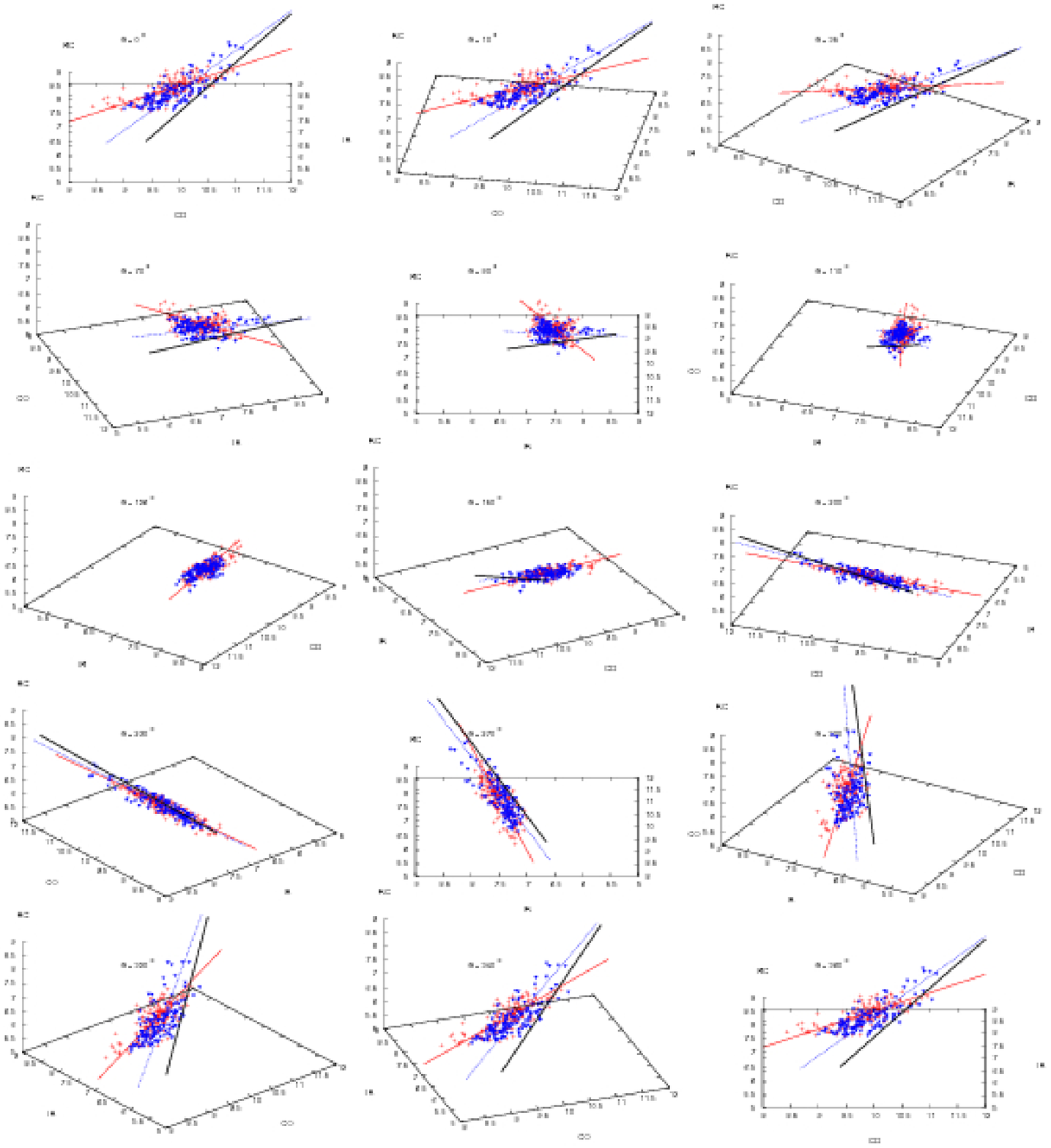}
\caption[]{{\bf RC, CO and 8 \,$\mu$m 3D correlation at 2\arcsec.} 
The x, y, and z axis are log(CO), log(IR) (8 $\mu$m), and log(RC), respectively.
Data at $\sim$ 2\arcsec resolution of 
northern and southern region of \object{NGC\,3627} are represented by red crosses and 
blue triangles, respectively.  Red solid and blue dotted lines represent the 
fit of these two set of data. The 
bolded black line corresponds to the fit of  6\arcsec RC, CO, and 8 \,$\mu$m data 
of the overall galaxy \object{NGC\,3627}.

}
\label{3Dline_NS}
\end{figure*}

\subsection{The comparison with H$\alpha$ images and the determination 
of thermal radio flux}
\label{therm}

The two regions of \object{NGC\,3627} show very different stellar and dust distributions. 
Some works report on observations of HII regions in 
this galaxy (see \citeauthor{hodge74} \citeyear{hodge74} 
and \citeauthor{chemin03}
\citeyear{chemin03}).
In the north bar end, crowded with stellar objects, dust and HII regions  
have been the subject of various works focused on supernovae SN1989B and SN1973R 
hosted in \object{NGC\,3627} (e.g., \citeauthor{vandyk99} \citeyear{vandyk99}, 
\citeauthor{wells94} \citeyear{wells94}).

Figure \ref{optha} shows the comparison between the RC and CO emissions (contours) 
and the positions of brightest HII regions (crosses), as reported in \cite{hodge74}.
Some distinct sources 
of RC emission lie over the positions of HII regions.
Without a radio spectral index to verify the thermal nature of the radiation from
these sources, we can only tentatively identify the radio emission as originating in 
the HII regions. Indeed, 
superposing the north bar end RC and H$\alpha$ images, we found that 
the spatial distribution of the brightest H$\alpha$ knots is 
well matched by the RC peaks of emission.
Instead, the CO emission is located 
on the edges of HII regions, tracing the dust arc-like distribution.
         
In the south bar end region
the RC and CO emissions show an extended distribution similar to that of dust, without 
resolved peaks of emission. 
In the H$\alpha$ image, there are some distinct bright knots also  
 identified by \cite{hodge74} as HII regions,   
but they do not have a point-like RC-counterpart.

We used the H$_{\alpha}$ image to estimate the RC thermal 
component to evaluate its effect on the correlations.
The typical thermal fraction of radio emission at 1 GHz, 
measured from integrated spectra 
 of spiral galaxies is of the order of 0.10. In particular,  
for \object{NGC\,3627}, \citeauthor{niklas97a} \citeyear{niklas97a} 
determined a thermal fraction $<$ 0.09.
We estimated the thermal radio flux from the emission in 
the H$\alpha$ recombination 
line (see, e.g., \citeauthor{caplan86} \citeyear{caplan86} 
and \citeauthor{niklas97a} \citeyear{niklas97a}),
using the \cite{niklas97a} Eq. 2
in the following way:

\begin{eqnarray}
\Big(\frac{\rm S_{\rm th}}{\rm mJy}\Big)= 2.238 \cdot 10^9 \cdot 
\Big( \frac {\rm S_{\rm H\alpha}}{\rm erg\, s^{-1}\, cm^{-2}}\Big)
\cdot \Big(\frac{\rm T_e}{\rm K} \Big)^{0.42}\nonumber\\
\times ~ \Biggl\{ 
ln [\frac{0.04995}{\nu/GHz}]+1.5 \cdot ln \Big(\frac{\rm T_e}{\rm K} \Big)
\Biggr\}
\label{Thha}
\end{eqnarray}

\noindent where S$_{\rm th}$ is 
  the thermal radio flux density at frequency $\nu$=1.4 GHz,  assuming 
a thermal electron temperature $\rm T_e$ of 10$^4$ K and measuring the 
H$\alpha$ flux: S$_{ \rm H \alpha}$.
The mean value of the thermal flux, measured in 
the same grid of point-to-point analysis in the northern region, is 0.02 mJy.
This value rises to 0.08 mJy in correspondence with the brightest HII regions
(\ref{hodge74}, indicated as bold-face crosses in Fig. \ref{optha}).
In the northern region,   
we found that the thermal fraction ranges from 0.07 to 0.11.
Even if our measure 
underestimates the thermal emission, owing to absorption of 
the H$\alpha$ emission by dust, 
it is still likely sufficient for the purpose of roughly estimating  
 the thermal flux density and to ensure that the thermal contribution 
to the radio emission on the spatial scales analyzed 
is of order of 10\%, not significantly different from that measured on 
integrated spectra \citep{niklas97a}.

\begin{table}
\begin{tabular}{lccccccc}\hline\hline

\multicolumn{1}{c}{}&
\multicolumn{3}{c}{\bf 24\,$\mu$m-RC-CO}& &
\multicolumn{3}{c}{\bf 8\,$\mu$m-RC-CO}\\ \hline
\hline
name&$\phi$&$\theta$&rms &&$\phi$&$\theta$&rms\\
\hline
\\
 \multicolumn{8}{c}{ 6\arcsec correlations}\\ \hline
NGC\,3351&0.94&0.61&0.16 && 0.77&0.69&0.13\\
NGC\,3521&0.77&0.35&0.14 && 0.57 & 0.39& 0.11\\
{\bf \object{NGC\,3627}} &{\bf 0.95}&{\bf 0.60}&{\bf 0.23}&& {\bf 0.63} & {\bf 0.75}& {\bf 0.19}\\
NGC\,4826&0.92&0.74&0.15&& 0.76 & 0.82& 0.14\\
NGC\,5194&0.97&0.52&0.23&& 0.81 & 0.60 &0.20\\
NGC\,7331&0.87&0.36&0.10&& 0.73 & 0.38 & 0.09\\
All&0.92&0.60&0.25&& 0.65& 0.72&0.24\\
\hline
\\
\multicolumn{8}{c}{ 2\arcsec correlations}\\ \hline
{\bf \object{NGC\,3627} N}&  & &  & & {\bf 0.32} & {\bf 0.32}  & {\bf 0.19}\\
{\bf \object{NGC\,3627} S}& & &   && {\bf 0.58}  &{\bf 0.62}   &{\bf 0.23}\\
\hline\hline
\end{tabular}
\caption{3D fit coefficients. Observed data have been fitted with a line 
in three-dimensional space. 
$\theta$ and $\phi$ are the direction angles of the line, rms is the scatter of the 
data.}
\label{tab:tetafi}
\end{table}

\subsection{The 3D view}
\label{3D}

The RC-CO-IR correlations are usually studied separately as 
projection in different planes.
What we observe is indeed a correlation between 
all three emissions. This three-dimensional aspect  cannot show up 
in the usual 2-dimensional 
description. Hence, we present a new perspective, 
a three-dimensional representation, 
 with which to visualize and study the RC-CO-IR correlation. 
We performed a fit to the data in a three-dimensional space, 
placing log(I$_{\rm CO}$), log(I$_{\rm IR}$), and  log(I$_{\rm RC}$) in  
the x, y and z axis, respectively.
We computed the line fitting of the observed points, by minimizing the 
distance of these points from the line:
\begin{displaymath}
\left\{\begin{array}{l}
x=x_0+\rho \cdot cos{\theta}\, cos{\phi}\\
y=y_0+\rho \cdot cos{\theta}\, sin{\phi}\\
z=z_0+\rho \cdot sin{\theta}\\ 
\end{array}\right.
\end{displaymath}
where ($x_0$,$y_0$,$z_0$) is a generic point through which 
the line passes, $\theta$ and $\phi$ 
are its direction angles, and $\rho$ is a parameter.
As the fit procedures are completely different, 
the 3D fit coefficients cannot be easily related to the power law indexes found for  
the two-dimensional correlations.
However, our purpose is to introduce  a promising technique to visualize 
the ``unified'' correlation between the 
three emissions. The extension of this representation to a larger sample 
of galaxies could possibly give insight into the physical explanation of 
the correlation itself. 

We used the six galaxies studied in \cite{io06} as a reference sample for  
the three-dimensional correlation.
Figure \ref{3Dline_24m} 
shows the RC-CO-24\,$\mu$m,  
with the plotted line representing the fit to all the data.
We conducted the same analysis on RC-CO-8\,$\mu$m correlation, but for 
the sake of conciseness, this figure is not shown. 
Values for the angles  $\theta$ and $\phi$, and the rms, resulting from the fit, for each galaxy 
are reported in Table \ref{tab:tetafi}.

In these correlations, we deal with surface brightnesses, and thus 
they are unaffected by distance biases. 
The scatter of the global 3-D correlation is of 0.24 dex, i.e., a factor 1.7.
This is the scatter among different galaxies, whereas the scatter within 
each galaxy is even smaller (see Table \ref{tab:tetafi}). 

Figure \ref{3Dline_NS} shows the comparison between the 
three-dimensional 2\arcsec\,RC-CO-8\,$\mu$m correlation  
in the two observed regions and the 6\arcsec\, correlation on the entire galaxy.
Red boxes and blue triangles represent measures obtained in
the north and south bar ends, respectively. The fitted lines for each set of data 
are also plotted. The bold line is the result of the fit to the data 
of the entire galaxy at 6\arcsec.
The line fitting data of the south bar end region presents direction angles 
similar to those of the entire galaxy, and resembling to those of 
all galaxies. On the contrary, the line fitting the northern region data  
has lower direction angles, indicating a flatter correlation.

\section{Discussion}
\label{disc}
Studying the RC, CO, and 8 $\mu$m emissions from two regions of 
spiral galaxy \object{NGC\,3627}, we found that, despite some minor 
local variations,  
the spatial distributions of RC, CO, and 8 $\mu$m emissions match very well.
The phenomenological model (see \citeauthor{bicay90}, \citeyear{bicay90} 
and, more recently, \citeauthor{murphy06L}, \citeyear{murphy06L}) 
proposed that the radio images 
should be a smoothed version of the IR maps due to the diffusion 
of cosmic-ray electrons 
from the star-forming regions. The results obtained in our study may 
suggest a more complex picture.

The propagation of cosmic rays along magnetic field lines is, 
however, still matter of debate.
It is generally recognized that the cosmic rays propagating along the galactic 
magnetic field are continuously scattered and accelerated 
(see \citeauthor{melrose68}, \citeyear{melrose68}). 
Since the late 1970's, the proposed theory of diffusive shock acceleration 
has been developed and widely accepted. This theory proposes the statistical 
Fermi's acceleration 
mechanism as the responsible for the particles acceleration \citep{fermi49} and charges 
the slowing of such particles with reflections at strong 
shock waves. 
Without going into detail (see \citeauthor{strong07} \citeyear{strong07},   
for a recent review on the cosmic ray propagation and 
interactions in the Galaxy) we can say that 
the bulk propagation of cosmic rays is slowed down to the velocity 
of magneto-hydrodynamic (MHD) waves, 
which is  the Alfv\'en speed, $\rm v_{\rm A}=\sqrt{\rm B^{2}/8 \pi \rho}= 
1.6\cdot 10^5\,n^{-1/2}\, \rm B_{\mu G}$, in cm s$^{-1}$, 
where B is the magnetic field strength, and $\rho$ and n are the density  
and number density of the gas, respectively.
For a typical disk magnetic field strength B$\sim\,5\,\mu$G \citep{condon92} 
and a density of the warm 
phase of the galactic disk ISM  (which may be responsible for 
turbulent diffusion of cosmic rays) n=0.3 cm$^{-3}$
\citep{turbbook}, 
the Alfv\'en velocity is $\sim$15 km s$^{-1}$.
The synchrotron lifetime $\tau$, at $\nu_{\rm c}$=1 GHz, is of 
about $10^8$ yrs. In these conditions, the diffusion scale-length of 
the radiating electrons, 
$\rm L_{\rm diff}\sim \rm v_{\rm A} \cdot \tau$, is of 1.5 kpc.

\begin{figure}[htp]
\hspace{-0.5cm}
\includegraphics[width=9 cm, bb=11 494 565 760]{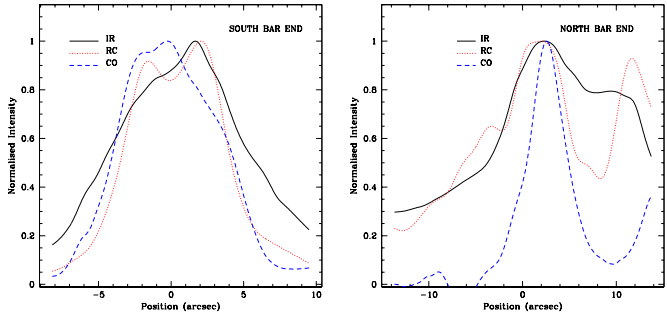}
\caption[]{Emission profiles of RC, CO and IR. {\it Left panel}: Profiles along 
the horizontal direction in the south bar end region. {\it Right panel}: Profiles 
along a slice in the thin arm. See Fig. \ref{optha} for the slice position. 
}
\label{prof}
\end{figure}

The scale we have examined (100 pc) is 
much lower than the electron diffusion scale. Nevertheless, in 
the two observed regions the RC emission does 
not appear to be more diffuse than that of CO and 8\,$\mu$m.
We compare the relative shapes of the emission profiles taking a 
one-dimensional slice of the RC, CO, and IR images.
Due to the clumpy distribution of the emissions at this resolution, 
the profiles obtained along 
some paths contain multiple emission maxima corresponding to resolved features. 
In Fig. \ref{prof}, the profiles  
obtained along two slices in the observed regions (the positions of slices 
are shown as black lines in Fig. \ref{optha}) 
are reported as an example.
The IR, RC, and CO profiles have been plotted with normalized peak amplitude 
to facilitate a comparison of their shapes and their spatial coincidence.
The slice in the southern region is taken along the horizontal direction.
 The RC and CO widths are comparable and not greater, as possibly expected,
 than the analogous width of the dust emission 
at 8\,$\mu$m. For the northern region we report the profiles 
taken along a slice in the thin arm structure, where all the three emission 
are detected. In this case the most concentrated emission appears to be the CO one,  
while the RC emission profile does not appear to be a smoothed version of the IR one.
In both regions, there is evidence that RC emission is not more diffuse than the 
CO and 8 $\mu$m emission. This is a surprising result, as the observed regions 
are much smaller than the expected electron diffusion scale.

It is important to stress that we found 
that the RC emission is well correlated to the emissions at $\mu$ m wavelengths 
in the northern and southern region. 
The $\mu$m emission traces a different dust component than that traced by the FIR 
emission. In practic, while the latter represents cold dust,
as mentioned before, the 4.5  $\mu$m emission contains component of hot dust, 
whereas in 5.8 and 8\,$\mu$m images the emission from PAH becomes significant.
The CO emission varies over three order of magnitude while dust emission spans only 
one order of magnitude.
In our opinion, this result
could be explained considering that the CO emission traces molecular gas where 
star formation 
may just begin, whereas the bulk of the 8\,$\mu$m emission, for example,
 comes from PAH molecules and these particles are in turn concentrated near regions 
of current star formation.

As we noted in previous sections, the north bar end region presents 
flat correlations of CO emission with both RC and IR. 
The flat RC-CO slope can be produced either by an higher RC brightness or a lower CO
integrated intensity with respect to the average RC-CO relation.
Assuming that the excess of RC brightness  
is due to the RC thermal component, the thermal fraction would be of 0.55.
This value is higher than what we measure from 
the H$_{\alpha}$ emission (see Sect. \ref{therm}), and thus this 
hypothesis is unlikely.
Otherwise, if the CO is lower than average, a possible explanation can be  
the photo-destruction 
of molecular clouds near HII regions by ultraviolet photons (for a review on 
photo-dissociation regions see, e.g., \citeauthor{PDRreview99} 
\citeyear{PDRreview99}). This option   can explain the lack of  CO 
emission in correspondence with bright HII regions with RC counterpart, 
nevertheless it still need to be quantitatively verified.
Certainly, more complex mechanisms are taking place in the observed 
regions, especially since the ends of the bar
might suffer from turbulent motions that influence the velocity field.

Since the RC brightness critically depends on the galaxy's magnetic field B
and cosmic-ray electron distribution N$_0$, and the separation 
between these two contributions  
is not yet clear, the interpretation of the results obtained is difficult. 
Spatially resolved spectral index studies at low frequencies, 
where the thermal emission 
component is seemingly negligible, 
can help to clarify the situation.

\section{Summary and concluding remarks}

We studied the correlations between the 1.4 GHz radio continuum, the CO,  
 and the IR emission at short-wavelengths in 
two regions of the spiral galaxy \object{NGC\,3627}.
We carried out new VLA and PdBI observations down to an angular 
resolution of order of 2\arcsec, which corresponds to a resolution of $\sim$ 100 pc 
at the distance of 11 Mpc.
We used the released IRAC and  
H$\alpha$ SINGS images of comparable resolution. 
Even if the scale we examined is much lower than the electron 
diffusion scale we do not 
find evidence of a diffusion of the RC emission with respect to the 
scales of CO and 8 $\mu$m, 
as would be expected by previous phenomenological models.

The observed regions are located at the northern and southern 
ends of the bar of \object{NGC\,3627} 
where two broad peaks of emission are shown both in RC and CO images. 
Our new high resolution observations permit a distinction 
between compact and extended emission that would not be possible with previous data. 
In particular, in the northern region we can resolve some RC peaks 
have a spatial coincidence with HII knots. 

We point out that our results refer to one single object,
and, in particular, to two distinct regions in this object, and we do not 
attempt to provide a definite answer to the question.    
Our main result is that in these two regions,
even at scales of 100 pc, where we would expect 
a break down of the correlations due to the cosmic rays diffusion,
 the RC emission is still correlated to that of CO.
Furthermore we found that the RC emission is also correlated to 
the 4.5, 5.8 and 8\,$\mu$m  dust emission.
Whatever these emissions trace, their correlation with RC emission on 
such a small scales is surprising and requires explaination. 
The study of a significant sample 
of galaxies would answer some of the questions left open.
 
Interestingly, in the northern region the RC-CO and IR-CO correlations 
have flatter slopes with respect to the average correlations,
 which might be due, among other options, to 
the photo-dissociation of molecular clouds near HII regions.
Despite this minor deviation, the spatial distributions of RC  follows well those of CO 
and IR,  tracing thin arc-like structures.
The widths of several structures, measured along slices across 
them, are similar in the RC, CO, and 8 $\mu$m maps. 
This indicates that the scale of RC emission is not larger than those 
of 8 $\mu$m and CO.

Furthermore, to give a complete description of the RC-CO, CO-mid IR, 
and RC-mid IR correlations, 
we have proposed a more concise and effective three-dimensional 
representation of the correlation, 
producing a three-dimensional fit of the observed data. We found  
that RC, CO, and mid-IR observed data are fitted by a line in 
 three-dimensional space with a scatter of about 0.25 dex. In our 
opinion, this is a promising technique to visualize the 
``unified'' correlation between the 
three emissions. 

This study is focused on two individual regions of the galaxy \object{NGC\,3627}, and 
the analysis of more objects has certainly to be done. Nevertheless,   
the observed correlation between RC, CO, and mid-IR emissions 
down to spatial scale of 100 pc 
may suggest that the coupling between electrons diffusion, losses, 
and injection is far more complex than expected.
Future works on spectral index at high resolution will 
help to separate the contributions to the RC brightness due to the strength of 
magnetic field from that due to the density of relativistic electrons.

\section*{Appendix A. CO(2-1) observations}

\begin{figure*}[ht!p]
\begin{center}
\includegraphics[angle=270, width=0.75\textwidth]{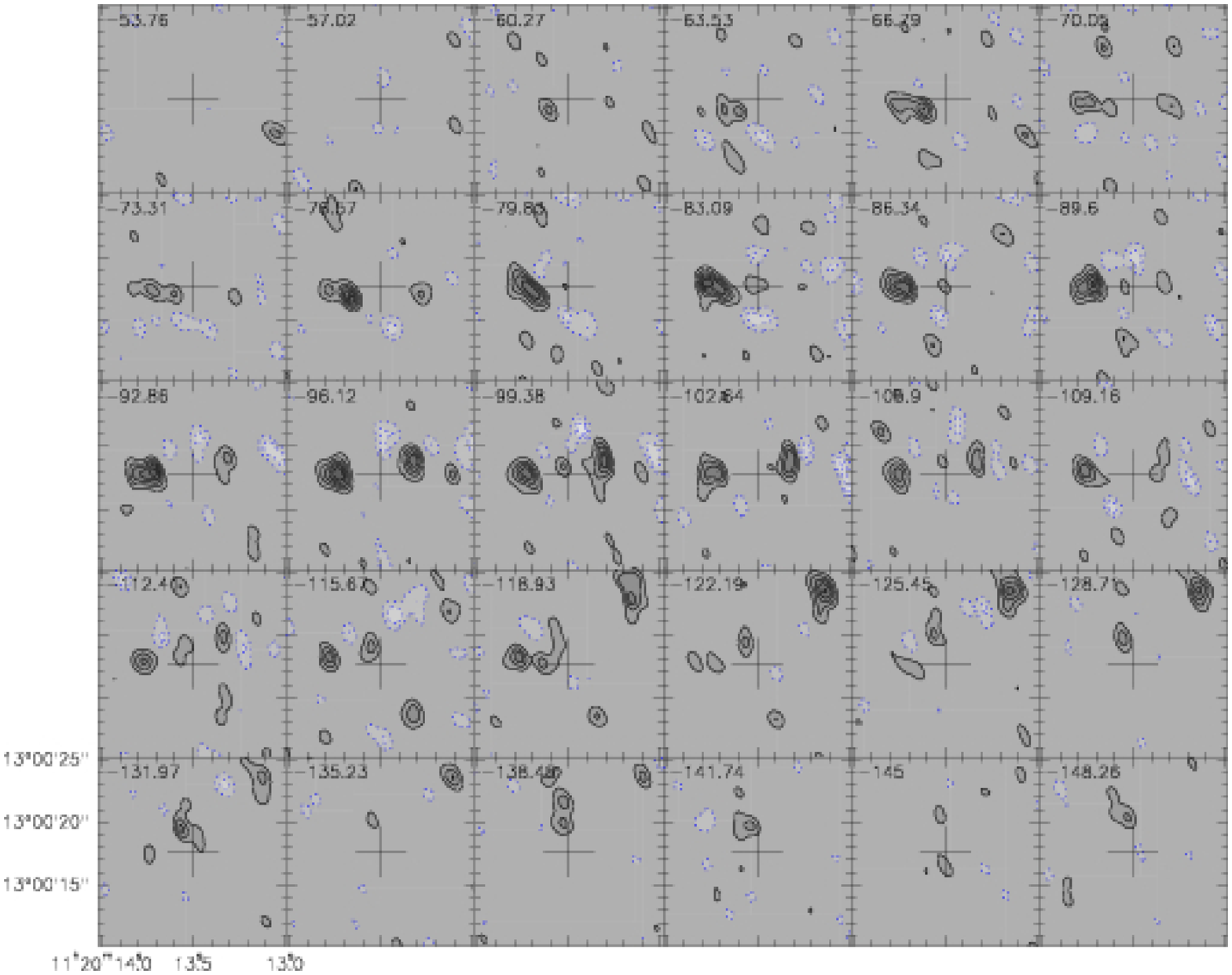}
\end{center}

\hfill
\begin{center}
\includegraphics[height=8 cm, bb=0 247 586 836]{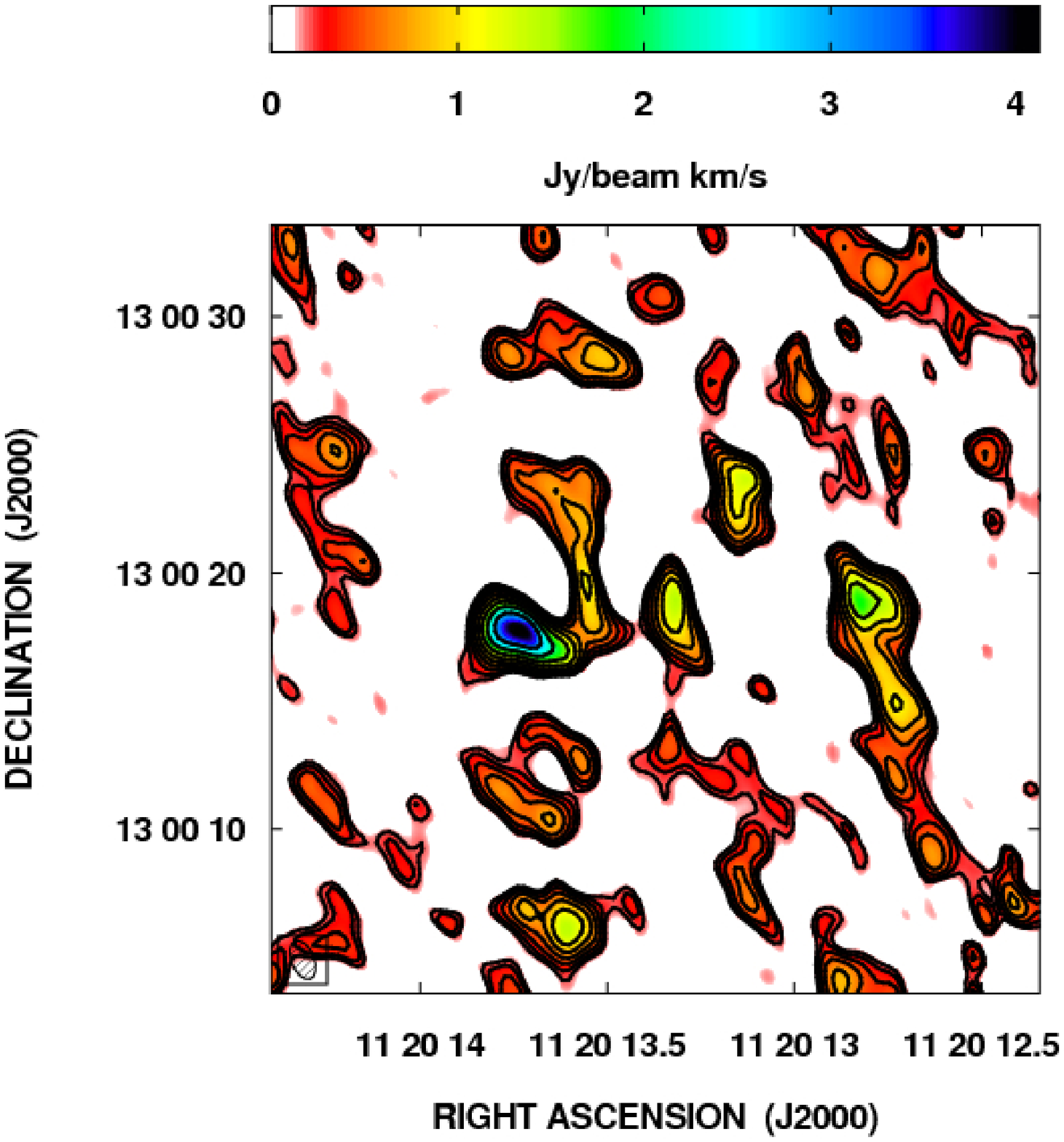}
\end{center}

\caption[]{{\bf North bar end of \object{NGC\,3627}.}
{\it Upper panel}: $^{12}$CO(2-1) velocity-channel maps observed with the PdBI.  
The spatial resolution  is of  
1\,\farcs59$\times$0\,\farcs88 at PA=22\deg. 
We map an area of view of 15\arcsec. The phase 
tracking center is indicated by a cross 
at $\alpha_{J2000}$=11$^h$20$^m$13.5$^s$ and 
  $\delta_{J2000}$=13\deg 00\arcmin 17\,\farcs7. Velocity-channels are 
displayed from v=$-$53.76 km s$^{-1}$ to v=$-$148.26 km s$^{-1}$ in steps of 
3.26  km s$^{-1}$. Velocities are referred to the LSR scale and the zero velocity corresponds to v$_0$=712.6 
km s$^{-1}$. Contour levels start from $-$20 mJy beam$^{-1}$ with step of 
20 mJy beam$^{-1}$.The rms noise is 3.4 mJy beam$^{-1}$ and only regions whose brightness is larger  
than 6-$\sigma$ are shown.

{\it Lower panel}: $^{12}$CO(2-1) integrated intensity.
Contours level start from 0.18 Jy km s$^{-1}$ beam$^{-1}$ and scale 
by a factor of $\sqrt{2}$. }
\label{CO21_nord}
\end{figure*}

\begin{figure*}[ht!p]
\begin{center}
\includegraphics[angle=270, width=0.75\textwidth]{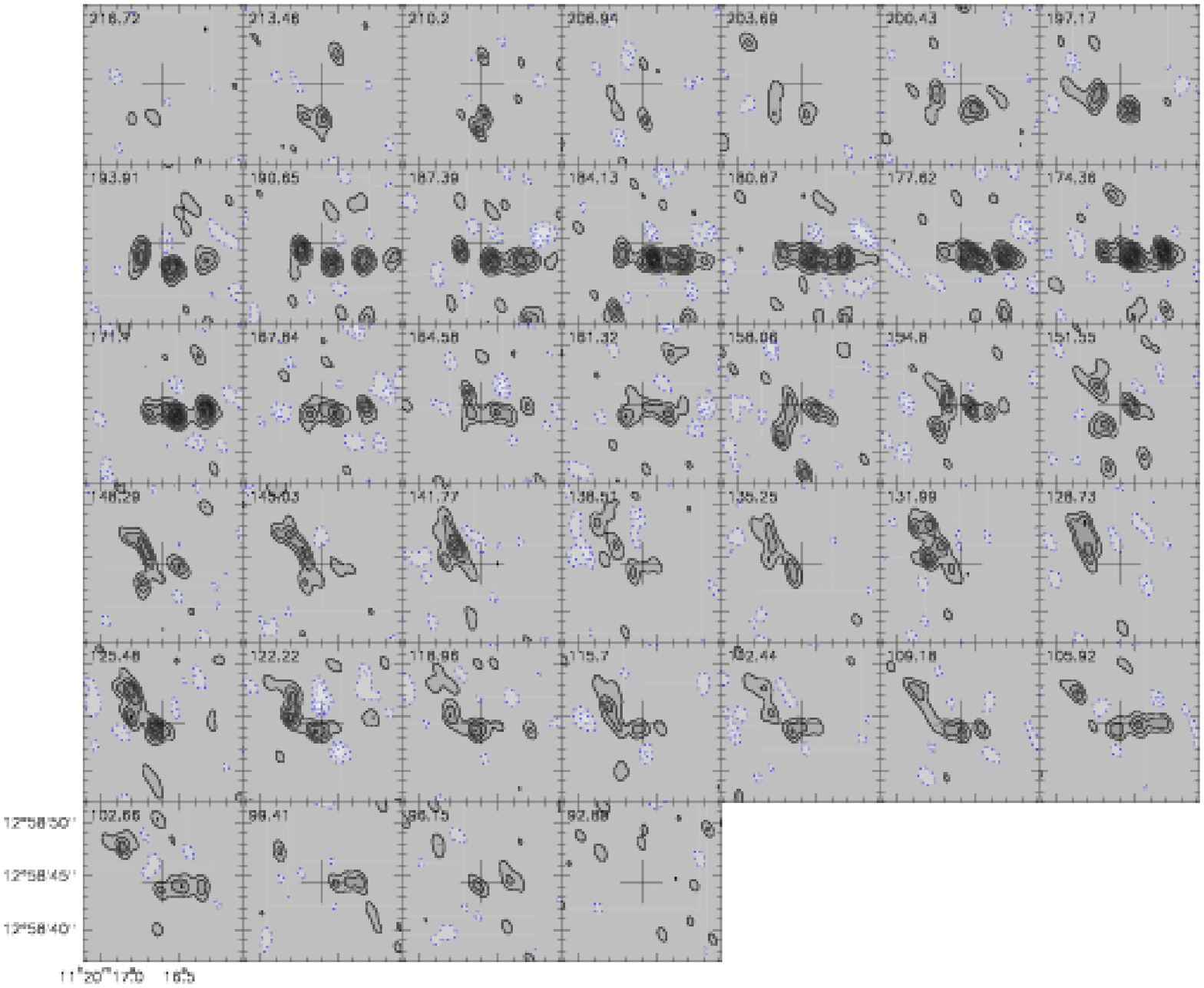}
\end{center}

\hfill
\begin{center}
\includegraphics[height=8 cm, bb=0 247 586 836]{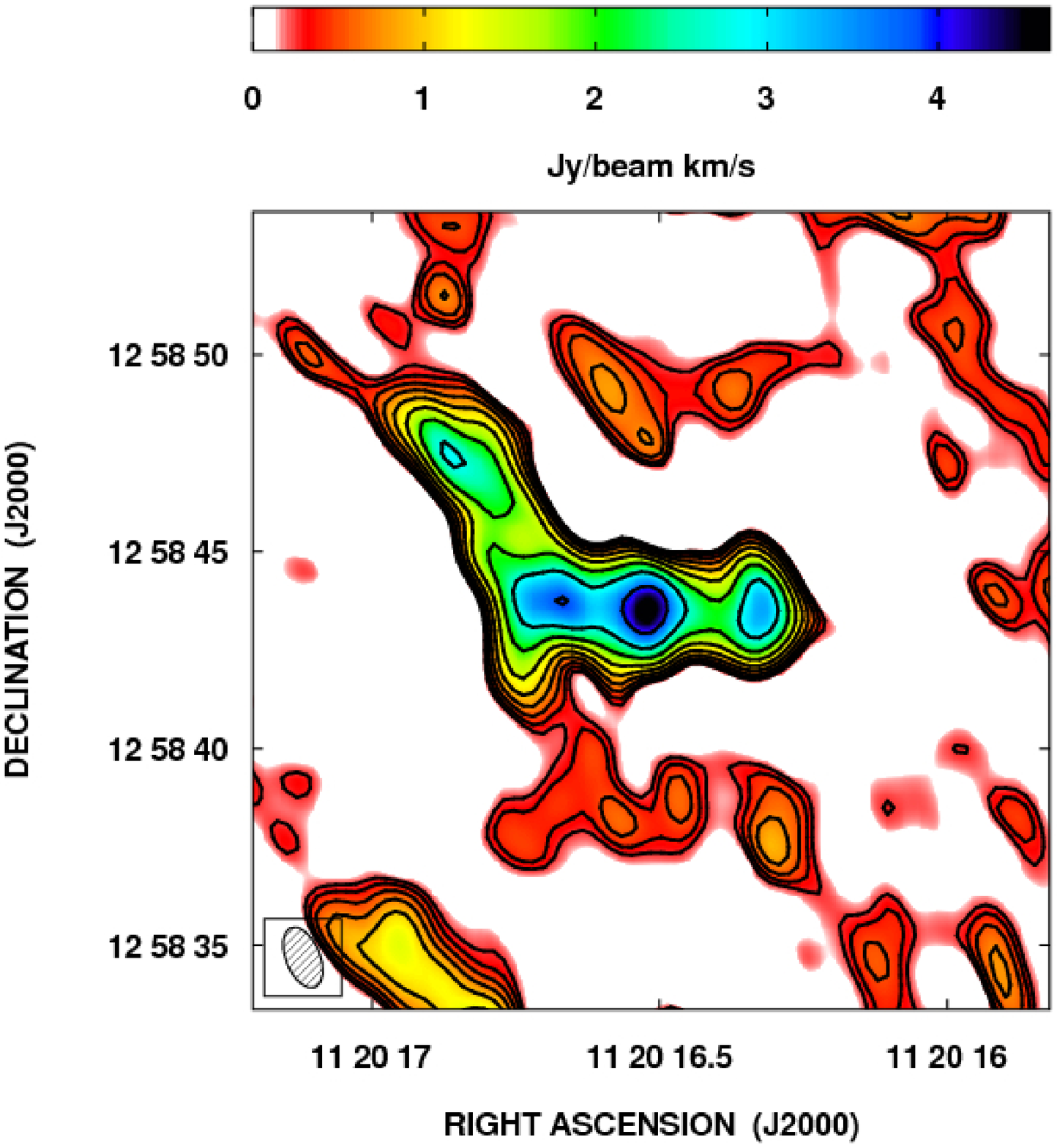}
\end{center}

\caption[]{{\bf South bar end of \object{NGC\,3627}.}
{\it Upper panel}: $^{12}$CO(2-1) velocity-channel maps observed with the PdBI. 
The spatial resolution is of  
1\,\farcs63$\times$0\,\farcs87 at PA=31\deg. 
We show a field of view of 15\arcsec. The phase 
tracking center is indicated by a cross 
at $\alpha_{J2000}$=11$^h$20$^m$ 16.6$^s$ and 
  $\delta_{J2000}$=12\deg 58\arcmin 44\,\farcs5. Velocity-channels are 
displayed from v=216.72 km s$^{-1}$ to v=92.89 km s$^{-1}$ in steps of 
3.26  km s$^{-1}$. Velocities are referred to the LSR and the zero velocity corresponds to v$_0$=712.6
km s$^{-1}$. Contour levels start from $-$20 mJy beam$^{-1}$ with step of 
20 mJy beam$^{-1}$. The rms noise is of 4 mJy beam$^{-1}$ and only regions whose brightness is larger 
than 5-$\sigma$ are shown. 

{\it Lower panel}: South bar end of \object{NGC\,3627}:$^{12}$CO(2-1) integrated intensity.
Contours level start from 0.24 Jy km s$^{-1}$ beam$^{-1}$ and scale 
by a factor of $\sqrt{2}$. }
\label{CO21_sud}
\end{figure*}

Figures \ref{CO21_nord}  and \ref{CO21_sud} show the 
$^{12}$CO(2-1) velocity-channel maps and the integrated intensity 
observed with the PdBI in the north and south bar end regions, respectively.

The $^{12}$CO(2-1) and  $^{12}$CO(1-0) emissions are detected in 
almost the same velocity 
range. The 230 GHz channels have been compressed to have the 
 velocity resolution of 3.26 Km s$^{-1}$.

Due to the particularly severe missing flux problem in these images, 
they are not reliable for 
quantitative studies. Nevertheless, we note that the northern $^{12}$CO(2-1) 
emission has a peak coincident with the $^{12}$CO(1-0) peak, and that in the 
southern region three different peaks of emission are resolved (Figure \ref{CO21_sud}).

\begin{acknowledgements}
This work employed extensive use of the NASA Extragalactic Database. 
This work has benefited from research funding from the European Community's 
Sixth Framework Programme.
The National Radio Astronomy Observatory is
operated by Associated Universities, Inc.,
under contract with the National Science
 Foundation. R.P. wishes to thank Philippe Salom\'e and IRAM staff for help 
provided during the PdBI observations and data reduction. 
R.P. also wishes to thank 
 Tamara Helfer and Michael Regan  for providing the BIMA data. 
We wish to thank an anonymous referee for providing a critical reading of 
the manuscript and for useful comments.
\end{acknowledgements}

\bibliography{N3627_hr}
\bibliographystyle{aa}

\end{document}